\documentclass[12pt,a4paper]{article}
\usepackage{cite}
\usepackage{fullpage}
\usepackage[centertags]{amsmath}
\allowdisplaybreaks[4]
\usepackage{amssymb}
\usepackage{bm}
\usepackage[dvips]{graphicx}
\usepackage{url}
\usepackage[dvips,hyperindex]{hyperref}
\begin{document}

\begin{flushright}
\begin{tabular}{r}
\textsf{2 December 2009}
\end{tabular}
\end{flushright}
\vspace{1cm}
\begin{center}
\Large\bfseries
Matter Effects in Active-Sterile Solar Neutrino Oscillations
\\[0.5cm]
\large\normalfont
C. Giunti\ensuremath{^{(a)}}, Y.F. Li\ensuremath{^{(a,b)}}
\\[0.5cm]
\normalsize\itshape
\setlength{\tabcolsep}{1pt}
\begin{tabular}{cl}
\ensuremath{(a)}
&
INFN, Sezione di Torino,
Via P. Giuria 1, I--10125 Torino, Italy
\\[0.3cm]
\ensuremath{(b)}
&
Department of Modern Physics, University of Science and
\\
&
Technology of China, Hefei, Anhui 230026, China
\end{tabular}
\end{center}
\begin{abstract}
The matter effects for solar neutrino oscillations are studied
in a general scheme with an arbitrary number of sterile neutrinos,
without any constraint on the mixing, assuming only a realistic hierarchy of neutrino squared-mass differences
in which the smallest squared-mass difference is effective in solar neutrino oscillations.
The validity of the analytic results are illustrated with a numerical solution of the evolution equation
in three examples of the possible mixing matrix in the simplest case of four-neutrino mixing.
\end{abstract}

\newpage

\section{Introduction}
\label{001}
\nopagebreak

The possible existence of sterile neutrinos
\cite{Pontecorvo:1968fh}
is one of the most interesting open questions in neutrino physics
(see Refs.~\cite{hep-ph/9812360,hep-ph/0202058,hep-ph/0405172,hep-ph/0506083,hep-ph/0606054,Giunti-Kim-2007,0704.1800}),
which may give important information on the physics beyond the Standard Model
(see Refs.~\cite{hep-ph/0111326,hep-ph/0603118}).
Sterile neutrinos are neutral fermions which
do not interact weakly,
as the three ordinary active neutrinos $\nu_{e}$, $\nu_{\mu}$, $\nu_{\tau}$ do.
Moreover,
sterile neutrinos can mix with active neutrinos,
leading to active-sterile neutrino oscillations which are observable
either through the disappearance of active neutrinos
or through flavor transitions generated by a squared-mass difference $\Delta{m}^2$ which is larger than
the solar (SOL) and atmospheric (ATM) mass differences
(see Refs.~\cite{hep-ph/9812360,hep-ph/0202058,hep-ph/0405172,hep-ph/0506083,hep-ph/0606054,Giunti-Kim-2007,0704.1800})
\begin{align}
\Delta{m}^{2}_{\text{SOL}}
\simeq
\null & \null
8 \times 10^{-5} \, \text{eV}^{2}
\,,
\label{002}
\\
\Delta{m}^{2}_{\text{ATM}}
\simeq
\null & \null
2.5 \times 10^{-3} \, \text{eV}^{2}
\,.
\label{003}
\end{align}
In fact, the oscillation data show that $\nu_{e}$, $\nu_{\mu}$, $\nu_{\tau}$
are mainly mixed with three light neutrinos
$\nu_{1}$, $\nu_{2}$, $\nu_{3}$
with masses such that
\begin{align}
&
\Delta{m}^{2}_{\text{SOL}} = \Delta{m}^{2}_{21}
\,,
\label{002a}
\\
&
\Delta{m}^{2}_{\text{ATM}} = |\Delta{m}^{2}_{31}| \simeq |\Delta{m}^{2}_{32}|
\,,
\label{003a}
\end{align}
with
$ \Delta{m}^{2}_{kj} \equiv m_{k}^2 - m_{j}^2 $.
The order of magnitude of the absolute values of the three neutrino masses
$m_{1}$, $m_{2}$, $m_{3}$
is bound to be smaller than about 1 eV
by $\beta$-decay, neutrinoless double-$\beta$ decay and cosmological data
(see Refs.~\cite{hep-ph/0211462,hep-ph/0606054,Giunti-Kim-2007,0704.1800}).

The possible existence of sterile neutrinos has been discussed by many authors
in connection with the indication of $\bar\nu_{\mu}\to\bar\nu_{e}$ oscillations
with $ \Delta{m}^2 \gtrsim 0.1 \, \text{eV}^2 $
found in the LSND experiment \cite{hep-ex/0104049}.
Since this squared-mass difference is much larger than
$\Delta{m}^{2}_{\text{ATM}}$,
the LSND oscillations can only be explained with
mixing of more than three neutrinos,
leading to the unavoidable introduction of additional sterile neutrinos.

The LSND signal has not been seen in other experiments
and is currently disfavored by the negative results of the
KARMEN \cite{hep-ex/0203021}
and
MiniBooNE \cite{0812.2243}
experiments
\cite{0705.0107,0710.2985,0906.1997}.
On the other hand,
indications in favor of active-sterile neutrino oscillations
\cite{hep-ph/0610352,0707.4593,0711.4222,0902.1992}
come from
the anomalous ratio of measured and predicted ${}^{71}\text{Ge}$
observed in the Gallium radioactive source experiments
GALLEX
\cite{Anselmann:1995ar,Hampel:1998fc}
and SAGE
\cite{Abdurashitov:1996dp,hep-ph/9803418,nucl-ex/0512041,0901.2200}
and from the
MiniBooNE \cite{0812.2243} low-energy anomaly.
Moreover,
sterile neutrinos may constitute the cosmological dark matter
and
active-sterile oscillations may
explain the pulsar velocities
\cite{%
0706.0178
,%
0707.1495
,%
0710.5180
,%
0803.2735
,%
0812.3256
,%
0906.2802
,%
0906.2968
,%
0906.4117
}.
Since the standard cosmological scenario constrains the masses, mixing, and number of neutrinos
\cite{%
hep-ph/0305075
,%
hep-ph/0308083
,%
0706.4084
,%
0711.2450
,%
0808.3910
,%
0810.5133
,%
0812.2249
},
sterile neutrinos may be heralds of alternative cosmological models
\cite{%
0712.1816
,%
0805.4014
,%
0906.3322
}.

Bearing in mind that the existence of sterile neutrinos which are not related to the LSND anomaly
remains possible,
the quest for sterile neutrinos
is one of the best tools for the exploration of physics beyond the Standard Model.
A convenient way to reveal the existence of sterile neutrinos is the search for disappearance of
active neutrinos due to active-sterile oscillations
\cite{0907.3145,0907.5487}.
The explanation of
the
Gallium radioactive source experiments anomaly
and the MiniBooNE low-energy anomaly through
active-sterile neutrino oscillations requires the disappearance of electron neutrinos
into sterile states
\cite{hep-ph/0610352,0707.4593,0711.4222,0902.1992}.
If these transitions exist,
they contribute also to the disappearance of solar electron neutrinos.
This effect, which has not been studied in detail so far,
is the object of study of this paper.

The matter effects for solar neutrino oscillations
in a four-neutrino scheme
was studied in Ref.~\cite{hep-ph/9908513},
where it was assumed that the electron neutrino has non-negligible mixing with only two massive neutrinos,
$\nu_{1}$ and $\nu_{2}$,
which generate the solar squared-mass difference
$ \Delta{m}^{2}_{\text{SOL}} = \Delta{m}^{2}_{21} $.
In this approximation,
electron neutrinos cannot oscillate into sterile states in the short distances required
for the explanation of the Gallium radioactive source experiments anomaly
and the MiniBooNE low-energy anomaly.
Hence, a combined analysis of the
Gallium radioactive source experiments anomaly,
the MiniBooNE low-energy anomaly
and solar neutrino data
in terms of active-sterile oscillations
\cite{CG-ML-YFL-QYL-2009}
requires an extension of the study in Ref.~\cite{hep-ph/9908513}
to the general case in which there is no constraint on the mixing of the electron neutrino.

In this paper we study the matter effects for solar neutrino oscillations
in a general scheme with an arbitrary number of sterile neutrinos,
without any constraint on the mixing.
Hence,
we consider the general case of mixing of three active neutrino fields
($\nu_{e}$, $\nu_{\mu}$, $\nu_{\tau}$)
and $N_{s}$ sterile neutrino fields:
\begin{equation}
\nu_{\alpha L} = \sum_{k=1}^{N} U_{\alpha k} \nu_{kL}
\qquad
(\alpha=e,\mu,\tau,s_{1},\ldots,s_{N_{s}})
\,,
\label{005}
\end{equation}
where $N=3+N_{s}$,
$\nu_{k}$ is the field of a massive neutrino with mass $m_{k}$,
and the subscript $L$ denotes the left-handed chiral component.
The numbering of the massive neutrinos is chosen in order to satisfy Eqs.~(\ref{002a}) and (\ref{003a})
with the first three neutrino masses and we assume that the additional neutrino masses are much larger than
$m_{1}$, $m_{2}$, $m_{3}$,
in order to generate short-baseline oscillations which could explain the
Gallium radioactive source experiments anomaly
and/or the MiniBooNE low-energy anomaly
\cite{hep-ph/0610352,0707.4593,0711.4222,0902.1992}
and could be observed in future laboratory experiments
\cite{hep-ph/0609177,hep-ph/0611178,hep-ex/0701004,0704.0388,0705.0107,0706.1462,0707.2481,0907.3145,0907.5487}
and through
astrophysical measurements
\cite{%
0706.0399
,%
0709.1937
,%
0806.3029
,%
0810.4057
,%
0909.5410
}.
Hence, we have the following hierarchy of squared-mass differences:\footnote{
The different case of mixing of two active neutrinos and a sterile neutrino
with two small squared-mass differences,
both of which contribute to solar neutrino oscillations,
has been studied in Ref.~\cite{hep-ph/0307266}.
}
\begin{equation}
\Delta{m}^2_{21}
\ll
|\Delta{m}^2_{31}|
\ll
\Delta{m}^2_{k1}
\quad
\text{for}
\quad
k \geq 3
\,.
\label{hie}
\end{equation}

The $ N \times N $ unitary mixing matrix $U$ can be parameterized in terms of
$ 3 + 3 N_{s} $
mixing angles
(see Ref.~\cite{Giunti-Kim-2007}).
Three mixing angles mix the three active neutrinos among themselves and the other $ 3 N_{s} $
mixing angles mix the three active neutrinos with the $ N_{s} $ sterile neutrinos.
There are also
$ 1 + 2 N_{s} $ Dirac phases
which affect neutrino oscillations
and
$ 2 + N_{s} $ Majorana phases
which do not affect neutrino oscillations
\cite{Bilenky:1980cx,Doi:1980yb,Langacker:1986jv}.
However,
in this paper we do not need an explicit parameterization of $U$.
We derive the flavor transition probabilities of solar neutrinos
in terms of the values of the elements of the mixing matrix.

The plan of the paper is as follows.
Section~\ref{013} is the main one, where we derive the flavor transition probabilities.
In Section~\ref{Non-Adiabatic} we discuss the peculiarities of the
flavor transition probabilities in the extreme non-adiabatic limit.
In Section~\ref{091} we illustrate the validity of our results in the simplest case of four-neutrino mixing,
considering some examples in which we
compare the values of the analytic approximation of the flavor transition probabilities derived in Section~\ref{013}
with those obtained with a numerical solution of the flavor evolution equation.
Conclusions are drawn in Section~\ref{092}.

\section{Evolution of Neutrino Flavors}
\label{013}
\nopagebreak

Solar neutrinos are described by the state
\begin{equation}
| \nu(x) \rangle
=
\sum_{\alpha=e,\mu,\tau,s_{1},\ldots,s_{N_{s}}} \psi_{\alpha}(x) | \nu_{\alpha} \rangle
\,,
\label{014}
\end{equation}
where $x$ is the distance from the production point in the core of the sun and
\begin{equation}
\psi_{\alpha}(0)
=
\delta_{\alpha e}
\,,
\label{015}
\end{equation}
with the obvious normalization
$ \sum_{\alpha} |\psi_{\alpha}(x)|^2 = 1 $.

From the mixing relation (\ref{005}) between flavor and massive neutrino fields
and the structure of the weak charged-current Lagrangian,
\begin{equation}
\mathcal{L}_{\text{CC}}
=
-
\frac{ g }{ \sqrt{2} }
\sum_{\alpha=e,\mu,\tau}
\sum_{k=1}^{N}
\overline{\ell_{\alpha L}} \gamma^{\rho} U_{\alpha k} \nu_{kL} W_{\rho}^{\dagger}
+
\text{H.c.}
\,,
\label{008}
\end{equation}
it follows that in vacuum the flavor states $| \nu_{\alpha} \rangle$
which describe neutrinos created in charged-current weak interactions
are unitary superpositions of massive states
$| \nu^{\text{V}}_{k} \rangle$
which are the quanta of the corresponding massive neutrino fields $\nu_{k}$
(see Ref.~\cite{Giunti-Kim-2007}):
\begin{equation}
| \nu_{\alpha} \rangle
=
\sum_{k=1}^{N} U_{\alpha k}^{*} | \nu^{\text{V}}_{k} \rangle
\qquad
(\alpha=e,\mu,\tau,s_{1},\ldots,s_{N_{s}})
\,.
\label{005a}
\end{equation}

The evolution of the flavor transition amplitudes $\psi_{\alpha}(x)$ is given by the Mikheev-Smirnov-Wolfenstein (MSW) equation\footnote{
We omit a diagonal term $E+m_{1}^2/2E$
which generates an irrelevant common phase.
}
\cite{Wolfenstein:1978ue,Mikheev:1985gs}
(see Ref.~\cite{Giunti-Kim-2007})
\begin{equation}
i \frac{d}{dx} \Psi
=
\frac{1}{2E} \left( U \mathcal{M}^2 U^{\dagger} + \mathcal{A} \right)
\Psi
\,,
\label{016}
\end{equation}
where $E$ is the neutrino energy and
\begin{align}
\Psi
=
\null & \null
\left(
\psi_{e},\psi_{\mu},\psi_{\tau},\psi_{s_{1}},\ldots,\psi_{s_{N_{s}}}
\right)^T
\,,
\label{017}
\\
\mathcal{M}^2
=
\null & \null
\text{diag}\!\left(0,\Delta{m}^2_{21},\Delta{m}^2_{31},\Delta{m}^2_{41},\ldots,\Delta{m}^2_{N1}\right)
\,,
\label{018}
\\
\mathcal{A}
=
\null & \null
\text{diag}\!\left(A_{\text{CC}}+A_{\text{NC}},A_{\text{NC}},A_{\text{NC}},0,\ldots\right)
\,,
\label{019}
\end{align}
with
$ \Delta{m}^2_{kj} = m_{k}^2 - m_{j}^2 $
and
\begin{equation}
A_{\text{CC}}=2E V_{\text{CC}}
\,,
\qquad
A_{\text{NC}}=2E V_{\text{NC}}
\,.
\label{020}
\end{equation}
The charge-current and
neutral-current
matter potentials are given by
\begin{equation}
V_{\text{CC}}=\sqrt{2}G_{\text{F}}N_{e}
\simeq
7.63 \times 10^{-14}
\,
\frac{ N_{e} }{ N_{\text{A}} \, \text{cm}^{-3} }
\,
\text{eV}
\,,
\qquad
V_{\text{NC}}=-\frac{1}{2}\sqrt{2}G_{\text{F}}N_{n}
\,,
\label{021}
\end{equation}
where $G_{\text{F}}$ is the Fermi constant,
$N_{e}$ is the electron number density,
$N_{n}$ is the neutron number density,
and
$N_{\text{A}}$ is Avogadro's number.
It is convenient to use the standard definition of the electron fraction
\begin{equation}
Y_{e} = \frac{ N_{e} }{ N_{e}+N_{n} }
\,.
\label{022}
\end{equation}
In a neutral medium we have
\begin{equation}
N_{e}
=
\frac{ \rho }{ \text{g} }
\,
N_{\text{A}} Y_{e}
\,,
\qquad
N_{n}
=
\frac{ \rho }{ \text{g} }
\,
N_{\text{A}} \left( 1 - Y_{e} \right)
\,.
\label{023}
\end{equation}
where $\rho$ is the density.
From Eqs.~(\ref{021}) and (\ref{023}), we obtain
\begin{equation}
A_{\text{NC}}
=
R_{\text{NC}} A_{\text{CC}}
\,,
\quad
\text{with}
\quad
R_{\text{NC}}
=
- \frac{ 1 - Y_{e} }{ 2 Y_{e} }
\,.
\label{025}
\end{equation}

For solar densities we have
\begin{equation}
A_{\text{CC}}
\sim
|A_{\text{NC}}|
\sim
\Delta{m}^2_{21}
\ll
|\Delta{m}^2_{k1}|
\quad
\text{for}
\quad
k \geq 3
\,.
\label{026}
\end{equation}
In order to decouple the flavor transitions generated by
$
A_{\text{CC}}
\sim
|A_{\text{NC}}|
\sim
\Delta{m}^2_{21}
$
from those generated by the larger squared-mass differences,
it is useful to work in the vacuum mass basis
\begin{equation}
\Psi^{\text{V}}
=
\left(
\psi^{\text{V}}_{1},\ldots,\psi^{\text{V}}_{N}
\right)^T
=
U^{\dagger} \Psi
\,,
\label{027}
\end{equation}
which satisfies the evolution equation
\begin{equation}
i \frac{d}{dx} \Psi^{\text{V}}
=
\frac{1}{2E} \left( \mathcal{M}^2 + U^{\dagger} \mathcal{A} U \right)
\Psi^{\text{V}}
\,,
\label{028}
\end{equation}
In this basis, the solar neutrino state in Eq.~(\ref{014}) is given by
\begin{equation}
| \nu(x) \rangle
=
\sum_{k=1}^{N} \psi^{\text{V}}_{k}(x) | \nu^{\text{V}}_{k} \rangle
\,,
\label{014a}
\end{equation}

In the vacuum basis the inequality in Eq.~(\ref{026})
implies that the evolution of each of the amplitudes
$\psi^{\text{V}}_{3},\ldots,\psi^{\text{V}}_{N}$
is decoupled from the others,
since the coupling due to the matter effects is negligible.
Therefore, we have
\begin{equation}
\psi^{\text{V}}_{k}(x)
\simeq
\psi^{\text{V}}_{k}(0)
\,
\exp\!\left( - i \, \frac{ \Delta{m}^2_{k1} x }{ 2 E } \right)
\,,
\quad
\text{for}
\quad
k \geq 3
\,,
\label{029}
\end{equation}
with
\begin{equation}
\psi^{\text{V}}_{k}(0)
=
U_{ek}^{*}
\,,
\label{030}
\end{equation}
from Eqs.~(\ref{015}) and (\ref{027}).

On the other hand,
the evolution of
$\psi^{\text{V}}_{1}$ and $\psi^{\text{V}}_{2}$
is coupled by the matter effect.
In order to solve it, we consider the truncated 1-2 sector of Eq.~(\ref{028}):
\begin{equation}
i \frac{d}{dx}
\begin{pmatrix}
\psi^{\text{V}}_{1}
\\
\psi^{\text{V}}_{2}
\end{pmatrix}
=
\frac{1}{2E}
\begin{pmatrix}
\sum_{\alpha} |U_{\alpha1}|^2 \mathcal{A}_{\alpha}
&
\sum_{\alpha} U_{\alpha1}^{*} U_{\alpha2} \mathcal{A}_{\alpha}
\\
\sum_{\alpha} U_{\alpha1} U_{\alpha2}^{*} \mathcal{A}_{\alpha}
&
\Delta{m}^2_{21}
+
\sum_{\alpha} |U_{\alpha2}|^2 \mathcal{A}_{\alpha}
\end{pmatrix}
\begin{pmatrix}
\psi^{\text{V}}_{1}
\\
\psi^{\text{V}}_{2}
\end{pmatrix}
\,,
\label{031}
\end{equation}
where the sums over the flavor index $\alpha$ cover only the
active neutrinos, which have a matter potential in Eq.~(\ref{019}).
It is convenient to subtract the diagonal term
\begin{equation}
\frac{1}{4E}
\left(
\Delta{m}^2_{21}
+
\sum_{\alpha} \left( |U_{\alpha1}|^2 + |U_{\alpha2}|^2 \right) \mathcal{A}_{\alpha}
\right)
\,,
\label{032}
\end{equation}
which generates an irrelevant common phase.
The resulting effective evolution equation can be written as
\begin{equation}
i \frac{d}{dx}
\begin{pmatrix}
\psi^{\text{V}}_{1}
\\
\psi^{\text{V}}_{2}
\end{pmatrix}
=
\frac{1}{4E}
\begin{pmatrix}
-
\Delta{m}^2_{21}
+
A \cos 2 \xi
&
A \sin 2 \xi e^{i\varphi}
\\
A \sin 2 \xi e^{-i\varphi}
&
\Delta{m}^2_{21}
-
A \cos 2 \xi
\end{pmatrix}
\begin{pmatrix}
\psi^{\text{V}}_{1}
\\
\psi^{\text{V}}_{2}
\end{pmatrix}
\,,
\label{033}
\end{equation}
with
\begin{align}
A \cos 2 \xi
=
\null & \null
\sum_{\alpha} \left( |U_{\alpha1}|^2 - |U_{\alpha2}|^2 \right) \mathcal{A}_{\alpha}
=
A_{\text{CC}} X
\,,
\label{034}
\\
A \sin 2 \xi e^{i\varphi}
=
\null & \null
2 \sum_{\alpha} U_{\alpha1}^{*} U_{\alpha2} \mathcal{A}_{\alpha}
=
A_{\text{CC}} Y e^{i\varphi}
\,,
\label{035}
\end{align}
where
\begin{align}
X
=
\null & \null
|U_{e1}|^2 - |U_{e2}|^2 + R_{\text{NC}} \sum_{\alpha=e,\mu,\tau} \left( |U_{\alpha1}|^2 - |U_{\alpha2}|^2 \right)
\nonumber
\\
=
\null & \null
|U_{e1}|^2 - |U_{e2}|^2 - R_{\text{NC}} \sum_{i=1}^{N_{s}} \left( |U_{s_{i}1}|^2 - |U_{s_{i}2}|^2 \right)
\,,
\label{036}
\\
Y
=
\null & \null
2
\left|
U_{e1}^{*} U_{e2} + R_{\text{NC}} \sum_{\alpha=e,\mu,\tau} U_{\alpha1}^{*} U_{\alpha2}
\right|
\nonumber
\\
=
\null & \null
2
\left|
U_{e1}^{*} U_{e2} - R_{\text{NC}} \sum_{i=1}^{N_{s}} U_{s_{i}1}^{*} U_{s_{i}2}
\right|
\,.
\label{037}
\end{align}
In these expressions we used the unitarity of the mixing matrix in order to express
the neutral-current contribution either through the active neutrino elements of the mixing matrix
or through the sterile neutrino elements.

One can immediately notice that,
apart from the complex phase $e^{i\varphi}$,
the effective evolution equation (\ref{033})
has the same structure as the effective evolution equation in the vacuum mass basis for
$\nu_e$-$\nu_\mu$ or $\nu_e$-$\nu_\tau$ two-neutrino mixing
with the two-neutrino mixing angle replaced by the effective mixing angle $\xi$
and $A_{\text{CC}}$ replaced by $A$,
with
\begin{equation}
\tan 2 \xi = \frac{Y}{X}
\,,
\qquad
A = A_{\text{CC}} \sqrt{ X^2 + Y^2 }
\,,
\label{038}
\end{equation}
In fact, one can easily check that in the limit of $\nu_e$-$\nu_\mu$ or $\nu_e$-$\nu_\tau$ two-neutrino mixing,
$\xi$ coincides with the two-neutrino mixing angle and $A$ becomes $A_{\text{CC}}$.
Therefore,
the effective mixing angle $\xi$ plays the same role as the usual two-neutrino mixing angle
in the solution of the flavor evolution equation.
However,
one must keep in mind that in general $\xi$ is not a pure mixing angle given by a specific parameterization of the mixing matrix,
because it depends on the neutral-current to charged-current ratio $R_{\text{NC}}$.

For simplicity, we consider a medium with constant electron fraction $Y_{e}$,
which implies a constant neutral-current to charged-current ratio
$R_{\text{NC}}$.
In this case, the effective mixing angle $\xi$
and the phase $\varphi$ are constant.
Moreover,
we can use the phase freedom in the parameterization of the mixing matrix in order to
eliminate $\varphi$.
In fact, the flavor transition probabilities obtained from the MSW evolution equation (\ref{016})
are invariant under the phase transformations
\begin{equation}
U_{\alpha k} \to e^{i\varphi_{\alpha}} U_{\alpha k} e^{i\varphi_{k}}
\,.
\label{039}
\end{equation}
which gives, from the definition in Eq.~(\ref{035}),
\begin{equation}
A \sin 2 \xi e^{i\varphi} \to A \sin 2 \xi e^{i(\varphi+\varphi_{2}-\varphi_{1})}
\,.
\label{040}
\end{equation}
Therefore,
choosing $ \varphi_{1}-\varphi_{2} = \varphi $,
the evolution equation (\ref{033}) is simplified to
\begin{equation}
i \frac{d}{dx}
\begin{pmatrix}
\psi^{\text{V}}_{1}
\\
\psi^{\text{V}}_{2}
\end{pmatrix}
=
\frac{1}{4E}
\begin{pmatrix}
-
\Delta{m}^2_{21}
+
A \cos 2 \xi
&
A \sin 2 \xi
\\
A \sin 2 \xi
&
\Delta{m}^2_{21}
-
A \cos 2 \xi
\end{pmatrix}
\begin{pmatrix}
\psi^{\text{V}}_{1}
\\
\psi^{\text{V}}_{2}
\end{pmatrix}
\,,
\label{041}
\end{equation}
which has the same structure as the effective evolution equation in the vacuum mass basis for
$\nu_e$-$\nu_\mu$ or $\nu_e$-$\nu_\tau$ two-neutrino mixing.

\begin{figure}[t!]
\begin{center}
\includegraphics*[bb=24 147 578 709, width=0.6\textwidth]{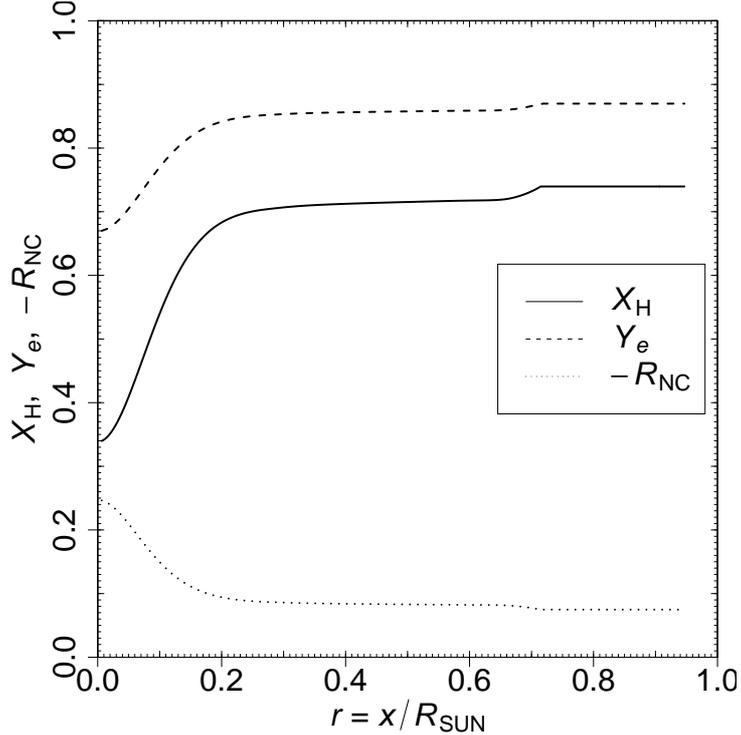}
\end{center}
\caption{ \label{fig01}
Hydrogen mass fraction $X_{\text{H}}$,
electron number fraction $Y_{e}=(1+X_{\text{H}})/2$,
and
neutral-current to charged-current ratio
$R_{\text{NC}}$
[Eq.~(\ref{025})]
in the BP04 Standard Solar Model
\cite{astro-ph/0402114}.
$ R_{\text{SUN}} \simeq 6.7 \times 10^{10} \, \text{cm} $
is the radius of the Sun.
}
\end{figure}

In practice,
$Y_{e}$ and $R_{\text{NC}}$ are not constant in the Sun,
as shown in Fig.~\ref{fig01}.
The electron fraction is almost constant in the
radiative
($ 0.25 \lesssim r \lesssim 0.7 $)
and
convective
($ r \gtrsim 0.7 $)
zones
where the Hydrogen/Helium mass fraction is practically equal to the primordial value.
In the core 
($ r \lesssim 0.25 $)
the electron fraction increases towards the center because of the accumulated
thermonuclear production of Helium during the 4.5 Gy of the life of the Sun.
However,
it is plausible that the effects on flavor transitions of the variation of $R_{\text{NC}}$
in the core is negligible if the transitions occur mainly in a resonance which is located in the radiative or in the convective zone.
Hence we solve the evolution equation
of neutrino flavors in the approximation of a constant $R_{\text{NC}}$.
In Section~\ref{091} we will show with numerical examples that the effects of the variation of $R_{\text{NC}}$
in the Sun are indeed negligible.

The effective Hamiltonian in
the evolution equation (\ref{041}) is diagonalized by the rotation
\begin{equation}
\begin{pmatrix}
\psi^{\text{V}}_{1}
\\
\psi^{\text{V}}_{2}
\end{pmatrix}
=
\begin{pmatrix}
\cos\omega
&
\sin\omega
\\
- \sin\omega
&
\cos\omega
\end{pmatrix}
\begin{pmatrix}
\psi^{\text{M}}_{1}
\\
\psi^{\text{M}}_{2}
\end{pmatrix}
\,,
\label{042}
\end{equation}
with
\begin{equation}
\tan2\omega
=
\frac{ A \sin 2 \xi }{ \Delta{m}^2_{21} - A \cos 2 \xi }
\,.
\label{043}
\end{equation}
The angle $\omega$
is the mixing angle between the vacuum mass basis and the effective mass basis in matter.
Its expression is similar to the one in
$\nu_e$-$\nu_\mu$ or $\nu_e$-$\nu_\tau$ two-neutrino mixing.
It is therefore useful to note that the evolution equation (\ref{041})
corresponds to the evolution of the effective interaction amplitudes
\begin{equation}
\begin{pmatrix}
\psi^{\text{I}}_{1}
\\
\psi^{\text{I}}_{2}
\end{pmatrix}
=
\begin{pmatrix}
\cos\xi
&
\sin\xi
\\
- \sin\xi
&
\cos\xi
\end{pmatrix}
\begin{pmatrix}
\psi^{\text{V}}_{1}
\\
\psi^{\text{V}}_{2}
\end{pmatrix}
\,,
\label{044}
\end{equation}
for which the matter potential is diagonal:
\begin{equation}
i \frac{d}{dx}
\begin{pmatrix}
\psi^{\text{I}}_{1}
\\
\psi^{\text{I}}_{2}
\end{pmatrix}
=
\frac{1}{4E}
\begin{pmatrix}
-
\Delta{m}^2_{21} \cos 2 \xi
+
A
&
\Delta{m}^2_{21} \sin 2 \xi
\\
\Delta{m}^2_{21} \sin 2 \xi
&
\Delta{m}^2_{21} \cos 2 \xi
-
A
\end{pmatrix}
\begin{pmatrix}
\psi^{\text{I}}_{1}
\\
\psi^{\text{I}}_{2}
\end{pmatrix}
\,.
\label{045}
\end{equation}
This basis of effective interaction amplitudes is physically important for the study of the flavor evolution,
because it is the basis in which the mixing generates transitions through the off-diagonal term
in the evolution equation.
Moreover,
in this basis one can see that there is a resonance when the diagonal terms are equal,
i.e. for $ A = A_{\text{R}} $,
with
\begin{equation}
A_{\text{R}} = \Delta{m}^2_{21} \cos 2 \xi
\,.
\label{052}
\end{equation}
The relations between the effective mass and interaction bases in matter
and the flavor basis
are discussed in Appendix~\ref{appendix}.

The connection between the effective interaction amplitudes $\psi^{\text{I}}_{1}$ and $\psi^{\text{I}}_{2}$
and the amplitudes $\psi^{\text{M}}_{1}$ and $\psi^{\text{M}}_{2}$ of the effective energy eigenstates in matter
is similar to that in two-neutrino mixing:
\begin{equation}
\begin{pmatrix}
\psi^{\text{I}}_{1}
\\
\psi^{\text{I}}_{2}
\end{pmatrix}
=
\begin{pmatrix}
\cos\xi_{\text{M}}
&
\sin\xi_{\text{M}}
\\
- \sin\xi_{\text{M}}
&
\cos\xi_{\text{M}}
\end{pmatrix}
\begin{pmatrix}
\psi^{\text{M}}_{1}
\\
\psi^{\text{M}}_{2}
\end{pmatrix}
\,,
\label{049}
\end{equation}
where $\xi_{\text{M}}$ is the effective mixing angle in matter given by
\begin{equation}
\xi_{\text{M}} = \xi + \omega
\,.
\label{050}
\end{equation}
From Eq.~(\ref{043}) we obtain
\begin{equation}
\tan2\xi_{\text{M}}
=
\frac{ \tan 2 \xi }{ 1 - \left( A / \Delta{m}^2_{21} \cos 2 \xi \right) }
\,.
\label{051}
\end{equation}
The effective mixing is maximal ($ \xi_{\text{M}} = \pi/4 $) in the resonance $ A = A_{\text{R}} $.
Notice that Eq.~(\ref{051}) and the resonance condition (\ref{052})
are similar to the corresponding equations in
$\nu_e$-$\nu_\mu$ or $\nu_e$-$\nu_\tau$ two-neutrino mixing
(see Ref.~\cite{Giunti-Kim-2007}),
with the two-neutrino mixing angle replaced by the effective mixing angle $\xi$
and $A_{\text{CC}}$ replaced by $A$.

The interesting limiting cases for the effective mixing angle in matter are:
\begin{alignat}{3}
\null & \null
\xi_{\text{M}} \simeq \xi
\,,
\quad
\null & \null
\null & \null
\omega \simeq 0
\,,
\quad
\null & \null
\null & \null
\text{for}
\quad
A \ll A_{\text{R}}
\,,
\label{053}
\\
\null & \null
\xi_{\text{M}} \simeq \frac{\pi}{2}
\,,
\quad
\null & \null
\null & \null
\omega \simeq \frac{\pi}{2} - \xi
\,,
\quad
\null & \null
\null & \null
\text{for}
\quad
A \gg A_{\text{R}}
\,.
\label{054}
\end{alignat}

The evolution of the amplitudes $\psi^{\text{M}}_{1}$ and $\psi^{\text{M}}_{2}$ of the effective energy eigenstates in matter
is given by
\begin{equation}
i \frac{d}{dx}
\begin{pmatrix}
\psi^{\text{M}}_{1}
\\
\psi^{\text{M}}_{2}
\end{pmatrix}
=
\frac{1}{4E}
\begin{pmatrix}
- \Delta{m}^2_{M}
&
- 4 i E \text{d} \omega / \text{d} x
\\
4 i E \text{d} \omega / \text{d} x
&
\Delta{m}^2_{M}
\end{pmatrix}
\begin{pmatrix}
\psi^{\text{M}}_{1}
\\
\psi^{\text{M}}_{2}
\end{pmatrix}
\,,
\label{055}
\end{equation}
with the effective squared-mass difference
\begin{equation}
\Delta{m}^2_{M}
=
\sqrt{
\left( \Delta{m}^2_{21} \cos 2 \xi - A \right)^2
+
\left( \Delta{m}^2_{21} \sin 2 \xi \right)^2
}
\,.
\label{056}
\end{equation}
The non-adiabatic off-diagonal transitions depend on
\begin{equation}
\frac{ \text{d} \omega }{ \text{d} x }
=
\frac{ \Delta{m}^2_{21} A \sin 2 \xi }{ 2 \left( \Delta{m}^2_{M} \right)^2 }
\,
\frac{ \text{d} \ln N_{e} }{ \text{d} x }
\,.
\label{057}
\end{equation}
It is convenient to define the adiabaticity parameter
\begin{equation}
\gamma
=
\frac{ \Delta{m}^2_{M} }{ 4 E | \text{d} \omega / \text{d} x | }
=
\frac{ \left( \Delta{m}^2_{M} \right)^3 }{ 2 E \Delta{m}^2_{21} A \sin 2 \xi | \text{d} \ln N_{e} / \text{d} x | }
\,.
\label{058}
\end{equation}
The evolution is adiabatic if $ \gamma \gg 1 $ all along the neutrino path.

The solution of the evolution equation (\ref{055}) leads to the averaged probabilities
\begin{align}
\overline{ |\psi^{\text{M}}_{1}(x)|^2 }
=
\null & \null
|\psi^{\text{M}}_{1}(0)|^2 \left( 1 - P_{12} \right)
+
|\psi^{\text{M}}_{2}(0)|^2 \, P_{12}
\,,
\label{059}
\\
\overline{ |\psi^{\text{M}}_{2}(x)|^2 }
=
\null & \null
|\psi^{\text{M}}_{1}(0)|^2 \, P_{12}
+
|\psi^{\text{M}}_{2}(0)|^2 \left( 1 - P_{12} \right)
\,,
\label{060}
\end{align}
where $P_{12}$
is the probability of non-adiabatic off-diagonal transitions in Eq.~(\ref{055}).
The averaging process washes out the interference between $\psi^{\text{M}}_{1}(x)$ and $\psi^{\text{M}}_{2}(x)$,
which is not measurable because of the energy resolution of the detector and the
uncertainty of the production region.
The initial values $\psi^{\text{M}}_{1}(0)$ and $\psi^{\text{M}}_{2}(0)$
are given by Eqs.~(\ref{030}) and (\ref{042}) with the effective mixing angle $\omega^{0}$ in the production region:
\begin{align}
\psi^{\text{M}}_{1}(0)
=
\null & \null
\cos\omega^{0} \, \psi^{\text{V}}_{1}(0) - \sin\omega^{0} \, \psi^{\text{V}}_{2}(0)
=
\cos\omega^{0} \, U_{e1}^{*} - \sin\omega^{0} \, U_{e2}^{*}
\,,
\label{061}
\\
\psi^{\text{M}}_{2}(0)
=
\null & \null
\sin\omega^{0} \, \psi^{\text{V}}_{1}(0) + \cos\omega^{0} \, \psi^{\text{V}}_{2}(0)
=
\sin\omega^{0} \, U_{e1}^{*} + \cos\omega^{0} \, U_{e2}^{*}
\,.
\label{062}
\end{align}
We can use the phase freedom in Eq.~(\ref{039})
in order to have real
$U_{e1}$ and $U_{e2}$,
keeping the reality of
$ 2 \sum_{\alpha} U_{\alpha1}^{*} U_{\alpha2} A_{\alpha} $ in Eq.~(\ref{035}),
since there are three arbitrary phases available:
$\varphi_{e}$,
$\varphi_{1}$, and
$\varphi_{2}$.
With this choice,
$U_{e1}$ and $U_{e2}$ can be written as
\begin{equation}
U_{e1} = \cos\vartheta_{e} \, \cos\chi_{e}
\,,
\quad
U_{e2} = \sin\vartheta_{e} \, \cos\chi_{e}
\,,
\quad
\text{with}
\quad
\sin^2\chi_{e} = \sum_{k=3}^{N} |U_{ek}|^2
\,.
\label{063}
\end{equation}
Then,
$\psi^{\text{M}}_{1}(0)$ and $\psi^{\text{M}}_{2}(0)$
are given by
\begin{equation}
\psi^{\text{M}}_{1}(0)
=
\cos\vartheta_{e}^{0} \cos\chi_{e}
\,,
\qquad
\psi^{\text{M}}_{2}(0)
=
\sin\vartheta_{e}^{0} \cos\chi_{e}
\,,
\label{064}
\end{equation}
with the effective mixing angle in the production region
\begin{equation}
\vartheta_{e}^{0} = \vartheta_{e} + \omega^{0}
\,.
\label{065}
\end{equation}
From Eq.~(\ref{086a}),
the electron neutrino state is given by
\begin{align}
| \nu_{e} \rangle
=
\null & \null
\left[
\cos(\vartheta_{e}+\omega) | \nu^{\text{M}}_{1} \rangle
+
\sin(\vartheta_{e}+\omega) | \nu^{\text{M}}_{2} \rangle
\right]
\cos\chi_{e}
+
\sum_{k=3}^{N} U_{\alpha k}^{*} | \nu^{\text{V}}_{k} \rangle
\nonumber
\\
=
\null & \null
\left[
\cos(\vartheta_{e}-\xi) | \nu^{\text{I}}_{1} \rangle
+
\sin(\vartheta_{e}-\xi) | \nu^{\text{I}}_{2} \rangle
\right]
\cos\chi_{e}
+
\sum_{k=3}^{N} U_{\alpha k}^{*} | \nu^{\text{V}}_{k} \rangle
\,.
\label{086b}
\end{align}
Hence,
an electron neutrino contains both
$ | \nu^{\text{I}}_{1} \rangle $
and
$ | \nu^{\text{I}}_{2} \rangle $
if
$\xi\neq\vartheta_{e}$.
In general, this is the case,
since the effective mixing angle $\xi$ in the effective evolution equation (\ref{033}) of the truncated 1-2 sector
is different from the electron neutrino mixing angle $\vartheta_{e}$
because of the neutral-current contribution to the matter potential
(see the discussion after Eq.~(\ref{077})).
In fact, as one can see from Eqs.~(\ref{036})--(\ref{038}),
the effective mixing angle $\xi$ does not depend only on the mixing of $\nu_{e}$,
but also on the mixings of $\nu_{\mu}$ and $\nu_{\tau}$
which affect the neutral-current contribution to the matter potential.

Since the matter effects in the detector are negligible,
the averaged flavor transition probabilities are given by
\begin{align}
\overline{P}_{\nu_{e}\to\nu_{\beta}}(x)
=
\null & \null
\overline{ |\psi_{\beta}(x)|^2 }
=
\overline{ \left| \sum_{k=1}^{N} U_{\beta k} \psi^{\text{V}}_{k}(x) \right|^2 }
\nonumber
\\
=
\null & \null
\overline{
\left|
\sum_{k=1}^{2} U_{\beta k} \psi^{\text{M}}_{k}(x)
+
\sum_{k=3}^{N} U_{\beta k} \psi^{\text{V}}_{k}(x)
\right|^2
}
\nonumber
\\
=
\null & \null
\sum_{k=1}^{2}
|U_{\beta k}|^2
\overline{ |\psi^{\text{M}}_{k}(x)|^2 }
+
\sum_{k=3}^{N} |U_{\beta k}|^2 |\psi^{\text{V}}_{k}(x)|^2
\nonumber
\\
=
\null & \null
\sum_{k=1}^{2}
|U_{\beta k}|^2
\overline{ |\psi^{\text{M}}_{k}(x)|^2 }
+
\sum_{k=3}^{N} |U_{\beta k}|^2 |U_{ek}|^2
\,,
\label{066}
\end{align}
where we have used Eqs.~(\ref{029}) and (\ref{030}).
Writing $|U_{\beta1}|^2$ and $|U_{\beta2}|^2$ as
\begin{equation}
|U_{\beta1}|^2 = \cos^2\vartheta_{\beta} \, \cos^2\chi_{\beta}
\,,
\quad
|U_{\beta2}|^2 = \sin^2\vartheta_{\beta} \, \cos^2\chi_{\beta}
\,,
\quad
\text{with}
\quad
\sin^2\chi_{\beta} = \sum_{k=3}^{N} |U_{\beta k}|^2
\,,
\label{067}
\end{equation}
and
using Eqs.~(\ref{059}), (\ref{060}) and (\ref{064}), we obtain
\begin{equation}
\overline{P}_{\nu_{e}\to\nu_{\beta}}
=
\left[
\frac{1}{2}
+
\left(
\frac{1}{2}
-
P_{12}
\right)
\cos2\vartheta_{\beta} \cos2\vartheta_{e}^{0}
\right]
\cos^2\chi_{\beta} \cos^2\chi_{e}
+
\sum_{k=3}^{N} |U_{\beta k}|^2 |U_{ek}|^2
\,.
\label{068}
\end{equation}
It is clear that in the limit of two-neutrino mixing,
in which $ \cos^2\chi_{e} = \cos^2\chi_{\beta} = 1 $,
the transition probability reduces to the well-known Parke formula \cite{Parke:1986jy}
(see Ref.~\cite{Giunti-Kim-2007}).

For neutrinos produced above the resonance,
where $A \gg A_{\text{R}}$,
from Eqs.~(\ref{054}) and (\ref{065}) we have
\begin{equation}
\vartheta_{e}^{0} \simeq \frac{\pi}{2} + \vartheta_{e} - \xi
\quad
\Longrightarrow
\quad
\cos2\vartheta_{e}^{0}
\simeq
- \cos2(\vartheta_{e}-\xi)
\,.
\label{069}
\end{equation}
In this case,
\begin{equation}
\overline{P}_{\nu_{e}\to\nu_{\beta}}
\simeq
\left[
\frac{1}{2}
-
\left(
\frac{1}{2}
-
P_{12}
\right)
\cos2\vartheta_{\beta} \cos2(\vartheta_{e}-\xi)
\right]
\cos^2\chi_{\beta} \cos^2\chi_{e}
+
\sum_{k=3}^{N} |U_{\beta k}|^2 |U_{ek}|^2
\,.
\label{070}
\end{equation}

The crossing probability $P_{12}$ can be inferred by using the similarity of
the evolution equation (\ref{055}) with that in two-neutrino mixing
\cite{Petcov:1988zj,Krastev:1988ci,Petcov:1988wv,Kuo:1989pn}
(see Ref.~\cite{Giunti-Kim-2007}):
\begin{equation}
P_{12}
=
\frac
{
\exp\left( - \frac{\pi}{2} \gamma_{\text{R}} F \right)
-
\exp\left( - \frac{\pi}{2} \gamma_{\text{R}} \frac{F}{\sin^{2}\xi} \right)
}
{
1
-
\exp\left( - \frac{\pi}{2} \gamma_{\text{R}} \frac{F}{\sin^{2}\xi} \right)
}
\,
\theta\!\left(A_{0}-A_{\text{R}}\right)
\,,
\label{071}
\end{equation}
where
$\gamma_{\text{R}}$
is the adiabaticity parameter at the resonance (\ref{052}),
which corresponds to the minimum of $\Delta{m}^2_{M}$:
\begin{equation}
\left( \Delta{m}^2_{M} \right)_{\text{R}} = \Delta{m}^2_{21} \sin 2 \xi
\,.
\label{072}
\end{equation}
Then, from Eq.~(\ref{058}), $\gamma_{\text{R}}$ is given by
\begin{equation}
\gamma_{\text{R}}
=
\frac
{\Delta{m}^{2} \sin^{2}2\xi}
{2 E \cos2\xi \left|\text{d}\ln N_{e}/\text{d}x\right|_{\text{R}}}
\,.
\label{073}
\end{equation}
The value of the parameter $F$ depends on the electron density
profile.
For an exponential density profile, which is a good
approximation for the solar neutrinos,
\cite{Petcov:1988zj,Krastev:1988ci,Petcov:1988wv,Pizzochero:1987fj,Toshev:1987jw,Kaneko:1987zz,Ito:1987vy,Kuo:1989pn,hep-ph/9712304}
\begin{equation}
F = 1 - \tan^2 \xi
\,.
\label{074}
\end{equation}
In Eq.~(\ref{071}) there is a $\theta$-factor which reduces $P_{12}$ to zero when
the neutrino is created at a density which is smaller than the resonance density
($A_{0}$ is the value of $A$ at production).

Let us finally notice that for solar neutrinos it is sufficient to calculate the $\nu_{e}$ survival probability
$ \overline{P}_{\nu_{e}\to\nu_{e}} $
and the sum of the transition probabilities into sterile neutrinos
$ \sum_{i=1}^{N_{s}} \overline{P}_{\nu_{e}\to\nu_{s_{i}}} $,
from which
$ \sum_{\beta=\mu,\tau} \overline{P}_{\nu_{e}\to\nu_{\beta}} $
is trivially obtained using the conservation of probability.
The separate transition probabilities into $\nu_{\mu}$ and $\nu_{\tau}$
are not needed because
$\nu_{\mu}$ and $\nu_{\tau}$ have the same interactions in the detector.
In fact,
$\nu_{\mu}$ and $\nu_{\tau}$ can be distinguished only through charged-current interactions
with production of the associated charged lepton,
which is forbidden by the low energy of solar neutrinos ($ E \lesssim 15 MeV $).
Hence,
$\nu_{\mu}$ and $\nu_{\tau}$ are detected only through neutral-current interactions
which are flavor-blind.

\section{Extreme Non-Adiabatic Limit}
\label{Non-Adiabatic}
\nopagebreak

It is interesting to note that in the extreme non-adiabatic limit
$ \gamma_{\text{R}} \ll 1 $
the transition probability $\overline{P}_{\nu_{e}\to\nu_{\beta}}$
is different from the averaged transition probability in vacuum (VAC)
\begin{align}
\overline{P}_{\nu_{e}\to\nu_{\beta}}^{\text{VAC}}
=
\null & \null
\sum_{k=1}^{N} |U_{\beta k}|^2 |U_{ek}|^2
\nonumber
\\
=
\null & \null
\left[
\frac{1}{2}
+
\frac{1}{2} \cos2\vartheta_{\beta} \cos2\vartheta_{e}
\right]
\cos^2\chi_{\beta} \cos^2\chi_{e}
+
\sum_{k=3}^{N} |U_{\beta k}|^2 |U_{ek}|^2
\,.
\label{075}
\end{align}
In fact,
since in the extreme non-adiabatic limit
\begin{equation}
P_{12}^{(\gamma_{\text{R}} \ll 1)} \simeq \cos^2 \xi
\,,
\label{076}
\end{equation}
from Eq.~(\ref{070})
we obtain
\begin{equation}
\overline{P}_{\nu_{e}\to\nu_{\beta}}^{(\gamma_{\text{R}} \ll 1)}
\simeq
\left[
\frac{1}{2}
+
\frac{1}{2} \cos2\vartheta_{\beta} \cos2(\vartheta_{e}-\xi) \cos2\xi
\right]
\cos^2\chi_{\beta} \cos^2\chi_{e}
+
\sum_{k=3}^{N} |U_{\beta k}|^2 |U_{ek}|^2
\,.
\label{077}
\end{equation}
This expression coincides with that in Eq.~(\ref{075}) only if $ \xi = \vartheta_{e} $.
From Eqs.(\ref{036})--(\ref{038})
one can see that $ \xi = \vartheta_{e} $ if
there is no neutral-current contribution in Eqs.(\ref{036}) and (\ref{037})
or
if the neutral-current contributions cancel in the ratio $Y/X$.
The first possibility is obviously realized in a neutron-free medium ($ R_{\text{NC}} = 0 $).
It is also realized if $ U_{s1} = U_{s2} = 0 $ for $ s = s_{1},\ldots,s_{N_{s}} $, which implies that
$ \sum_{\alpha=e,\mu,\tau} |U_{\alpha1}|^2 = \sum_{\alpha=e,\mu,\tau} |U_{\alpha2}|^2 = 1 $
and
$ \sum_{\alpha=e,\mu,\tau} U_{\alpha1}^{*} U_{\alpha2} = 0 $.
This is obviously the case in three-neutrino mixing (no sterile neutrinos).
The second possibility is realized in four-neutrino mixing with $ U_{e3} = U_{e4} = 0 $ \cite{hep-ph/9908513}:
in that case
\begin{equation}
U_{e1} = \cos\vartheta_{e}
\,,
\quad
U_{e2} = \sin\vartheta_{e}
\,,
\label{078}
\end{equation}
and
\begin{equation}
U_{s1} = - \sin\vartheta_{e} \cos\chi_{s}
\,,
\quad
U_{s2} = \cos\vartheta_{e} \cos\chi_{s}
\,.
\label{079}
\end{equation}
Using the unitarity of the mixing matrix, we have
\begin{align}
\null & \null
\sum_{\alpha=e,\mu,\tau} \left( |U_{\alpha1}|^2 - |U_{\alpha2}|^2 \right)
=
|U_{s2}|^2 - |U_{s1}|^2
=
\cos2\vartheta_{e} \cos^2\chi_{s}
\,,
\label{080}
\\
\null & \null
2 \sum_{\alpha=e,\mu,\tau} U_{\alpha1}^{*} U_{\alpha2}
=
-
U_{s1}^{*} U_{s2}
=
\sin2\vartheta_{e} \cos^2\chi_{s}
\,,
\label{081}
\end{align}
leading to
\begin{align}
X
=
\null & \null
\cos2\vartheta_{e} \left( 1 + R_{\text{NC}} \cos^2\chi_{s} \right)
\,,
\label{082}
\\
Y
=
\null & \null
\sin2\vartheta_{e} \left( 1 + R_{\text{NC}} \cos^2\chi_{s} \right)
\,,
\label{083}
\end{align}
and $\tan2\xi=X/Y=\tan2\vartheta_{e}$.

In the case of two-neutrino mixing the averaged flavor transition probability in the
extreme non-adiabatic case coincides with the averaged flavor transition probability in vacuum
because the flavor states are conserved in the resonance.
A $\nu_{e}$ created above the resonance is practically equal to a $\nu^{\text{M}}_{2}$,
which travels undisturbed to the resonance, where it is still a $\nu_{e}$
(it only develops a harmless phase).
If the crossing of the resonance is extremely non-adiabatic, the $\nu_{e}$ emerges unchanged
from the resonance
(apart from an irrelevant phase)
and the flavor transition measured on Earth is only due to the vacuum oscillations from the
Sun to the Earth.
On the other hand, in the case of mixing with sterile neutrinos, we have seen in Eq.~(\ref{086b})
that,
if
$\xi\neq\vartheta_{e}$, a $\nu_{e}$ contains both
$ \nu^{\text{I}}_{1} $
and
$ \nu^{\text{I}}_{2} $,
which are conserved if the crossing of the resonance is extremely non-adiabatic.
The different phases developed by
$ \nu^{\text{I}}_{1} $
and
$ \nu^{\text{I}}_{2} $
before and during resonance crossing
make the averaged flavor transition probability different from that in vacuum.
We can show that explicitly following the evolution of the
solar neutrino state
$ | \nu(x) \rangle $.
If the neutrino is created as a $\nu_{e}$ above the resonance,
the initial state is, from Eqs.~(\ref{054}), (\ref{063}), and (\ref{086a})
\begin{equation}
| \nu(0) \rangle
=
| \nu_{e} \rangle
=
\left[
-
\sin(\vartheta_{e}-\xi) | \nu^{\text{M}}_{1} \rangle
+
\cos(\vartheta_{e}-\xi) | \nu^{\text{M}}_{2} \rangle
\right]
\cos\chi_{e}
+
\sum_{k=3}^{N} U_{ek}^{*} | \nu^{\text{V}}_{k} \rangle
\,.
\label{090}
\end{equation}
Since from the production to the resonance the effective massive states develop different phases,
just before the resonance the solar neutrino state is
\begin{align}
| \nu(x_{\text{R}}^{-}) \rangle
=
\null & \null
\left[
-
\sin(\vartheta_{e}-\xi) e^{i\phi_{1}(x_{\text{R}}^{-})} | \nu^{\text{M}}_{1} \rangle
+
\cos(\vartheta_{e}-\xi) e^{i\phi_{2}(x_{\text{R}}^{-})} | \nu^{\text{M}}_{2} \rangle
\right]
\cos\chi_{e}
\nonumber
\\
\null & \null
+
\sum_{k=3}^{N} U_{ek}^{*} e^{i\phi_{k}(x_{\text{R}}^{-})} | \nu^{\text{V}}_{k} \rangle
\,.
\label{101}
\end{align}
Since
above the resonance
$ | \nu^{\text{I}}_{1} \rangle \simeq | \nu^{\text{M}}_{2} \rangle $
and
$ | \nu^{\text{I}}_{2} \rangle \simeq - | \nu^{\text{M}}_{1} \rangle $,
we have
\begin{align}
| \nu(x_{\text{R}}^{-}) \rangle
=
\null & \null
\left[
\sin(\vartheta_{e}-\xi) e^{i\phi_{1}(x_{\text{R}}^{-})} | \nu^{\text{I}}_{2} \rangle
+
\cos(\vartheta_{e}-\xi) e^{i\phi_{2}(x_{\text{R}}^{-})} | \nu^{\text{I}}_{1} \rangle
\right]
\cos\chi_{e}
\nonumber
\\
\null & \null
+
\sum_{k=3}^{N} U_{ek}^{*} e^{i\phi_{k}(x_{\text{R}}^{-})} | \nu^{\text{V}}_{k} \rangle
\,.
\label{102}
\end{align}
If the crossing of the resonance is extremely non-adiabatic,
there are no transitions among the effective interaction states
$| \nu^{\text{I}}_{1} \rangle$
and
$| \nu^{\text{I}}_{2} \rangle$,
which develop only different phases.
Hence,
just after the resonance we have
\begin{align}
| \nu(x_{\text{R}}^{+}) \rangle
=
\null & \null
\left[
\sin(\vartheta_{e}-\xi) e^{i\phi_{1}(x_{\text{R}}^{+})} | \nu^{\text{I}}_{2} \rangle
+
\cos(\vartheta_{e}-\xi) e^{i\phi_{2}(x_{\text{R}}^{+})} | \nu^{\text{I}}_{1} \rangle
\right]
\cos\chi_{e}
\nonumber
\\
\null & \null
+
\sum_{k=3}^{N} U_{ek}^{*} e^{i\phi_{k}(x_{\text{R}}^{+})} | \nu^{\text{V}}_{k} \rangle
\,.
\label{103}
\end{align}
Since after the resonance the neutrino propagates practically in vacuum,
it is convenient to express its state in terms of the propagating massive states:
\begin{align}
| \nu(x>x_{\text{R}}) \rangle
=
\null & \null
\left[
- \sin(\vartheta_{e}-\xi) \sin\xi e^{i\phi_{1}(x)}
+ \cos(\vartheta_{e}-\xi) \cos\xi e^{i\phi_{2}(x)}
\right]
\cos\chi_{e} | \nu^{\text{V}}_{1} \rangle
\nonumber
\\
\null & \null
+
\left[
\sin(\vartheta_{e}-\xi) \cos\xi e^{i\phi_{1}(x)}
+ \cos(\vartheta_{e}-\xi) \sin\xi e^{i\phi_{2}(x)}
\right]
\cos\chi_{e} | \nu^{\text{V}}_{2} \rangle
\nonumber
\\
\null & \null
+
\sum_{k=3}^{N} U_{ek}^{*} e^{i\phi_{k}(x)} | \nu^{\text{V}}_{k} \rangle
\,.
\label{103a}
\end{align}
Since
$ \phi_{1}(x) \neq \phi_{2}(x) $,
if
$\xi\neq\vartheta_{e}$
the absolute values of the amplitudes of
$| \nu^{\text{V}}_{1} \rangle$
and
$| \nu^{\text{V}}_{2} \rangle$
are different from those of
\begin{equation}
| \nu_{e} \rangle
=
\left[
\cos\vartheta_{e} | \nu^{\text{V}}_{1} \rangle
+
\sin\vartheta_{e} | \nu^{\text{V}}_{2} \rangle
\right]
\cos\chi_{e}
+
\sum_{k=3}^{N} U_{ek}^{*} | \nu^{\text{V}}_{k} \rangle
\,.
\label{104}
\end{equation}
Therefore,
if
$\xi\neq\vartheta_{e}$
the averaged probability
$ \overline{P}_{\nu_{e}\to\nu_{\beta}}^{(\gamma_{\text{R}} \ll 1)} $
cannot be equal to
$\overline{P}_{\nu_{e}\to\nu_{\beta}}^{\text{VAC}}$.
Instead, from Eq.~(\ref{104}) we obtain
\begin{align}
\overline{P}_{\nu_{e}\to\nu_{\beta}}^{(\gamma_{\text{R}} \ll 1)}
=
\null & \null
\overline{ | \langle \nu_{\beta} | \nu(x>x_{\text{R}}) \rangle |^2 }
\nonumber
\\
=
\null & \null
\Big\{
\cos^2\vartheta_{\beta}
\left[
\sin^2(\vartheta_{e}-\xi) \sin^2\xi
+
\cos^2(\vartheta_{e}-\xi) \cos^2\xi
\right]
\nonumber
\\
\null & \null
+
\sin^2\vartheta_{\beta}
\left[
\sin^2(\vartheta_{e}-\xi) \cos^2\xi
+
\cos^2(\vartheta_{e}-\xi) \sin^2\xi
\right]
\Big\}
\cos^2\chi_{e}
\cos^2\chi_{\beta}
\nonumber
\\
\null & \null
+
\sum_{k=3}^{N} |U_{\beta k}|^2 |U_{ek}|^2
\,.
\label{089}
\end{align}
After simple trigonometric transformations one obtains Eq.~(\ref{077}).

\section{Examples}
\label{091}
\nopagebreak

In this Section we
illustrate the validity of the flavor transition probability (\ref{068})
and the crossing probability (\ref{071})
by considering some examples in the simplest case of four-neutrino mixing.
In this case we have only one sterile neutrino
$ \nu_{s} \equiv \nu_{s_{1}} $.

\begin{table}[t!]
\begin{center}
\begin{tabular}{c|ccccc}
&
$U$
&
$Y_{e}$
&
$\tan^2\xi$
&
$\cos2(\vartheta_{e}-\xi)$
&
$\cos2\vartheta_{s}$
\\
\hline
\\
M1
&
$
\begin{pmatrix}
0.82 & 0.52 & 0.22 & 0.071
\\
\cdots & \cdots & \cdots & \cdots
\\
\cdots & \cdots & \cdots & \cdots
\\
-0.22 & 0.32 & -0.2 & 0.9
\end{pmatrix}
$
&
$
\begin{array}{c}
0.85 \\ \\ 0.1
\end{array}
$
&
$
\begin{array}{c}
0.4 \\ \\ 0.22
\end{array}
$
&
$
\begin{array}{c}
1 \\ \\ 0.97
\end{array}
$
&
-0.33
\\
\\
\hline
\\
M2
&
$
\begin{pmatrix}
0.82 & 0.52 & 0.22 & 0.1
\\
\cdots & \cdots & \cdots & \cdots
\\
\cdots & \cdots & \cdots & \cdots
\\
-0.22 & 0.1 & 0.16 & 0.96
\end{pmatrix}
$
&
$
\begin{array}{c}
0.85 \\ \\ 0.1
\end{array}
$
&
$
\begin{array}{c}
0.4 \\ \\ 0.2
\end{array}
$
&
$
\begin{array}{c}
1 \\ \\ 0.96
\end{array}
$
&
0.67
\\
\\
\hline
\\
M3
&
$
\begin{pmatrix}
0.8 & 0.51 & 0.22 & 0.22
\\
\cdots & \cdots & \cdots & \cdots
\\
\cdots & \cdots & \cdots & \cdots
\\
0.18 & 0.25 & -0.29 & -0.91
\end{pmatrix}
$
&
$
\begin{array}{c}
0.85 \\ \\ 0.1
\end{array}
$
&
$
\begin{array}{c}
0.41 \\ \\ 0.67
\end{array}
$
&
$
\begin{array}{c}
1 \\ \\ 0.97
\end{array}
$
&
-0.33
\\
\\
\hline
\hline
\end{tabular}
\end{center}
\caption{ \label{tab01}
Electron and sterile rows of the three examples of mixing matrices M1, M2, and M3
considered in the text.
For each of them we list the value of $\cos2\vartheta_{s}$
and the values of
$\tan^2\xi$
and
$\cos2(\vartheta_{e}-\xi)$
corresponding to two values of the electron fraction $Y_{e}$.
}
\end{table}

Since we have derived the flavor transition probability without using a specific parameterization of the mixing matrix,
we continue with this approach.
We consider the three examples of mixing matrices M1, M2, and M3 in Tab.~\ref{tab01}.
In all of them the value of $\vartheta_{e}$ corresponds to the best-fit value
of the solar mixing angle obtained in two- and three-neutrino mixing fits of solar and KamLAND data
\cite{hep-ph/0506083,hep-ph/0605195,0808.2016,0810.5443}:
\begin{equation}
\tan^2 \vartheta_{e} = \tan^2 \vartheta_{\text{SOL}} \simeq 0.4
\,.
\label{153}
\end{equation}
In Tab.~\ref{tab01} we give the electron and sterile rows of the mixing matrix
which contain all the information needed to calculate the probabilities
$\overline{P}_{\nu_{e}\to\nu_{e}}$
and
$\overline{P}_{\nu_{e}\to\nu_{s}}$,
from which it is immediate to obtain the remaining measurable probability
$\sum_{\beta=\mu,\tau}\overline{P}_{\nu_{e}\to\nu_{\beta}}$
as
$ 1 - \overline{P}_{\nu_{e}\to\nu_{e}} - \overline{P}_{\nu_{e}\to\nu_{s}} $.

In the first example we choose
\begin{equation}
\text{M1}:
\qquad
|U_{e3}|^2 = 0.05
\,, \quad
|U_{e4}|^2 = 0.005
\,, \quad
|U_{s1}|^2 = 0.05
\,, \quad
|U_{s2}|^2 = 0.1
\,.
\label{M1}
\end{equation}
We choose relatively large values of
$ |U_{s1}|^2 $ and $ |U_{s2}|^2 $
in order to have significant $\nu_{e}\to\nu_{s}$ transitions.
The values of $ |U_{e3}|^2 $ and $ |U_{e4}|^2 $
are sufficiently small to be compatible with the
reactor constraints
\cite{0711.2018,0711.4222}.
The value of $ |U_{e4}|^2 $
is constrained by the results of short-baseline (SBL) reactor experiments
in which the effective mixing angle is given by
$ \sin^2 2 \vartheta_{ee}^{\text{SBL}} = 4 |U_{e4}|^2 \left( 1 - |U_{e4}|^2 \right) $
(see Section~7.7 of Ref.~\cite{Giunti-Kim-2007}).
The Bugey limit
$ \sin^2 2 \vartheta_{ee}^{\text{SBL}} \lesssim 0.1 $
for
$ \Delta{m}^2_{41} \gtrsim 3 \times 10^{-2} \, \text{eV}^2 $
\cite{Declais:1995su}
implies that
\begin{equation}
|U_{e4}|^2 \lesssim 0.025
\,.
\label{ue4}
\end{equation}
Neglecting the small value of $ |U_{e4}|^2 $,
the effective mixing angle in long-baseline (LBL) reactor experiments is given by
$ \sin^2 2 \vartheta_{ee}^{\text{LBL}} = 4 |U_{e3}|^2 \left( 1 - |U_{e3}|^2 \right) $
(see Section~7.8 of Ref.~\cite{Giunti-Kim-2007}).
The Chooz limit
$ \sin^2 2 \vartheta_{ee}^{\text{LBL}} \lesssim 0.2 $
for
$ \Delta{m}^2_{31} = \Delta{m}^2_{\text{ATM}} \simeq 2.5 \times 10^{-3} \, \text{eV}^2 $
\cite{Apollonio:2002gd}
implies that \cite{hep-ph/9802201,hep-ph/0208026}
\begin{equation}
|U_{e3}|^2 \lesssim 0.05
\,.
\label{ue3}
\end{equation}
Therefore,
in the first example (M1) we choose a value of $|U_{e3}|^2$ at the level of the current experimental upper bound
and a value of
$|U_{e4}|^2$ much smaller that the current experimental upper bound.

The second example has been obtained by choosing
\begin{equation}
\text{M2}:
\qquad
|U_{e3}|^2 = 0.05
\,, \quad
|U_{e4}|^2 = 0.01
\,, \quad
|U_{s1}|^2 = 0.05
\,, \quad
|U_{s2}|^2 = 0.01
\,.
\label{M2}
\end{equation}
Hence, we consider the same values of
$|U_{e3}|^2$
and
$|U_{s1}|^2$
as in M1,
but the value of
$|U_{s2}|^2$
is much smaller,
and the value of
$|U_{e4}|^2$
is larger,
not far from the current experimental upper bound in Eq.~(\ref{ue4}).

In the third example we choose
\begin{equation}
\text{M3}:
\qquad
|U_{e3}|^2 = 0.05
\,, \quad
|U_{e4}|^2 = 0.05
\,, \quad
|U_{s1}|^2 = 0.03
\,, \quad
|U_{s2}|^2 = 0.06
\,.
\label{M3}
\end{equation}
The value of
$|U_{e4}|^2$ is larger than the current experimental upper bound in Eq.~(\ref{ue4}).
It is motivated by a possible explanation of
the
Gallium radioactive source experiments anomaly
and the MiniBooNE low-energy anomaly through
relatively large very-short-baseline active-sterile neutrino oscillations
\cite{0707.4593,0902.1992}.
This possibility requires a relaxing of the reactor constraint on $|U_{e4}|^2$
which could be due to
an underestimate of systematic uncertainties in the calculation of the reactor neutrino flux
or
to a violation of the CPT symmetry
\cite{0902.1992}.

\begin{figure}[p!]
\begin{center}
\begin{tabular}{cc}
%
\includegraphics*[bb=18 16 270 211, width=0.46\textwidth]{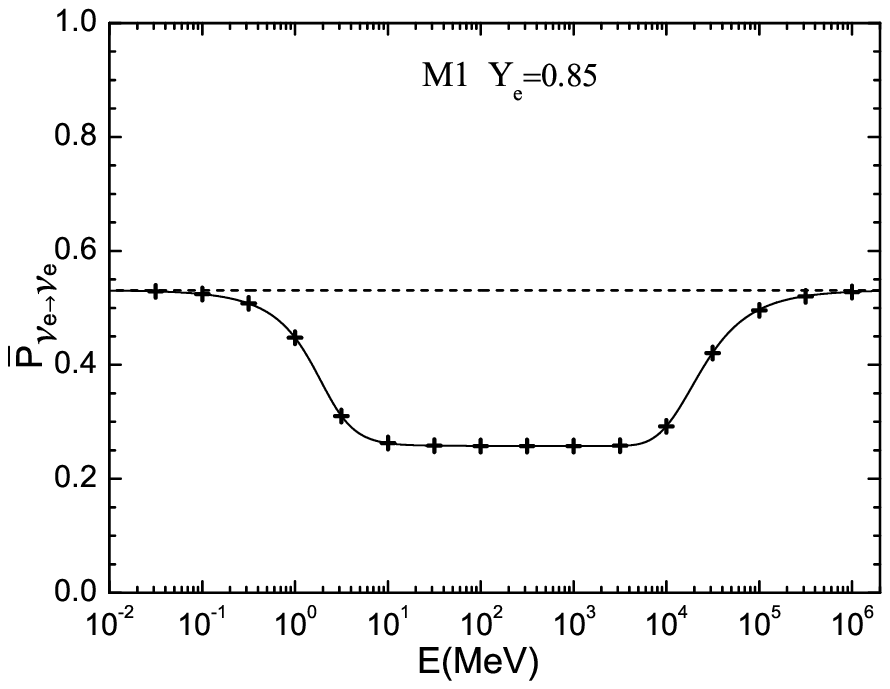}
&
\includegraphics*[bb=18 16 270 209, width=0.46\textwidth]{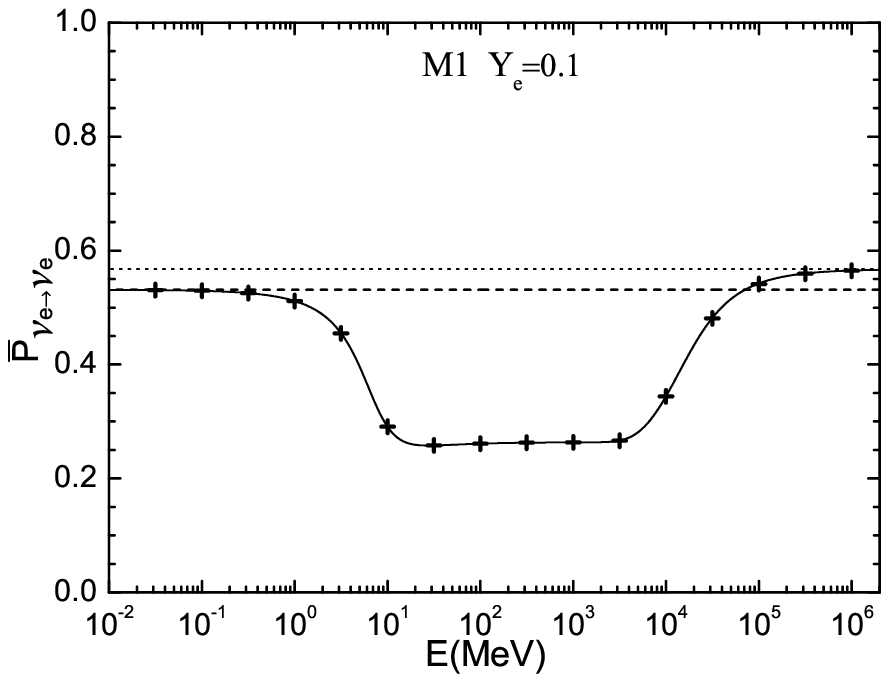}
\\
\includegraphics*[bb=18 16 270 209, width=0.46\textwidth]{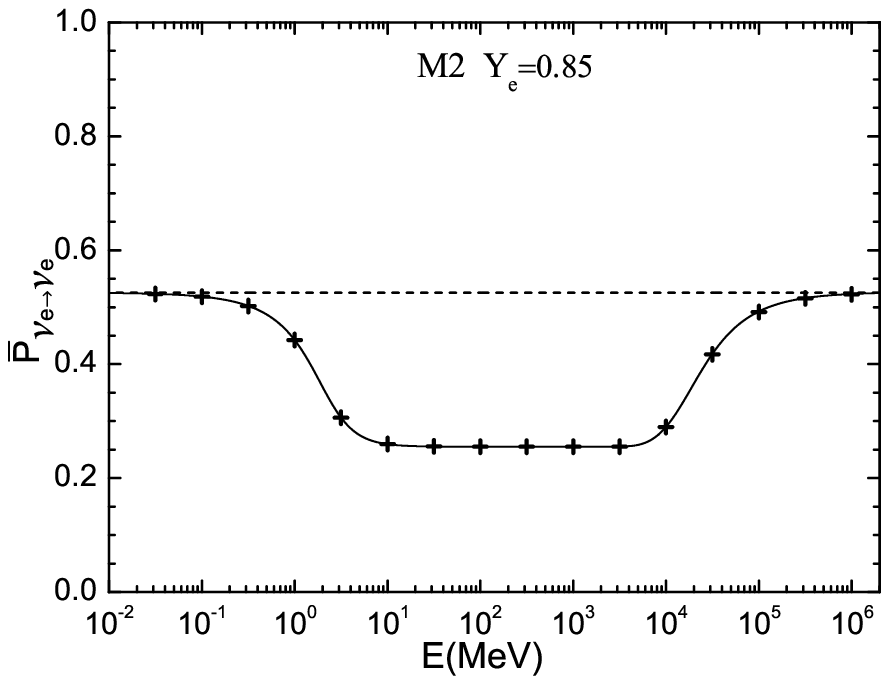}
&
\includegraphics*[bb=18 16 270 208, width=0.46\textwidth]{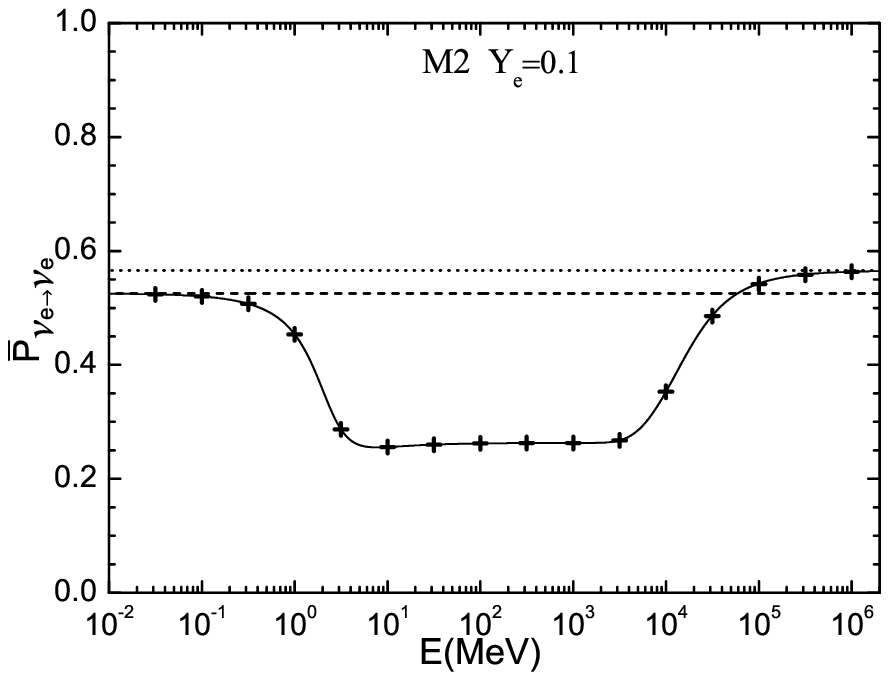}
\\
\includegraphics*[bb=18 16 270 208, width=0.46\textwidth]{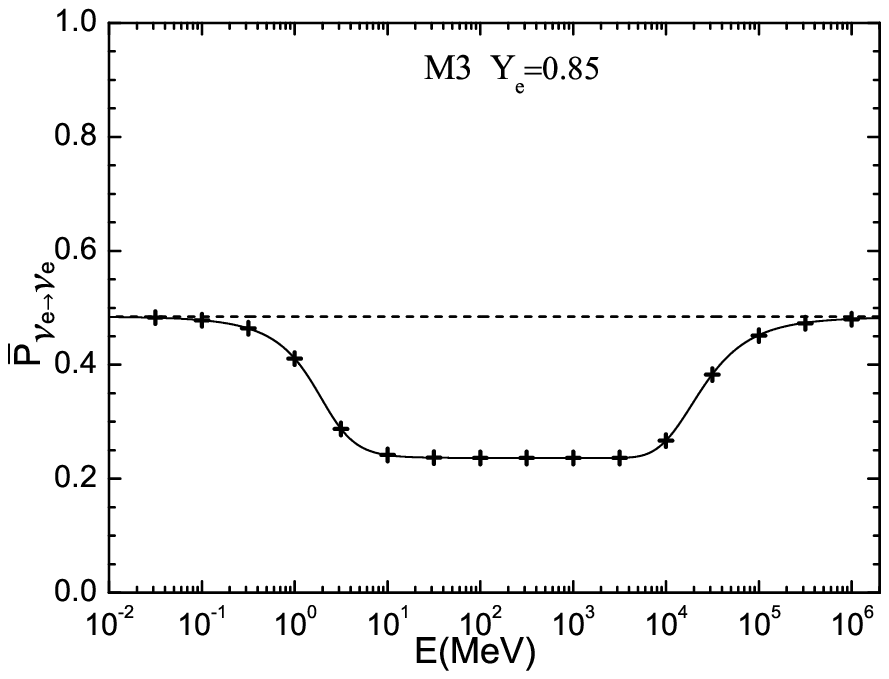}
&
\includegraphics*[bb=18 16 271 209, width=0.46\textwidth]{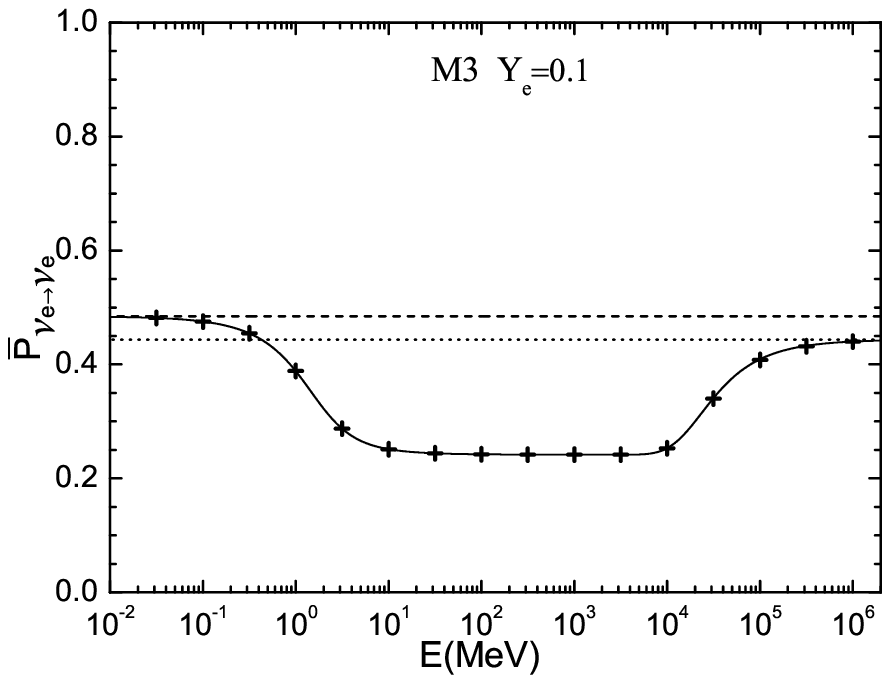}
\end{tabular}
\end{center}
\caption{ \label{fig02}
Averaged probability of $\nu_{e}$ survival as a function of the neutrino energy $E$
for the three examples of mixing matrices M1, M2, and M3 in Tab.~\ref{tab01}.
For each of them we consider the two values of the electron fraction $Y_{e}$ in Tab.~\ref{tab01}.
In each plot,
the solid line is obtained with the analytic expression in Eq.~(\ref{068}),
the points are obtained with a numerical solution of the evolution equation,
the horizontal dashed line shows the value of $\overline{P}_{\nu_{e}\to\nu_{e}}^{\text{VAC}}$,
and
the horizontal dotted line shows the value of $\overline{P}_{\nu_{e}\to\nu_{e}}^{(\gamma_{\text{R}} \ll 1)}$.
}
\end{figure}

\begin{figure}[p!]
\begin{center}
\begin{tabular}{cc}
%
\includegraphics*[bb=18 16 270 209, width=0.46\textwidth]{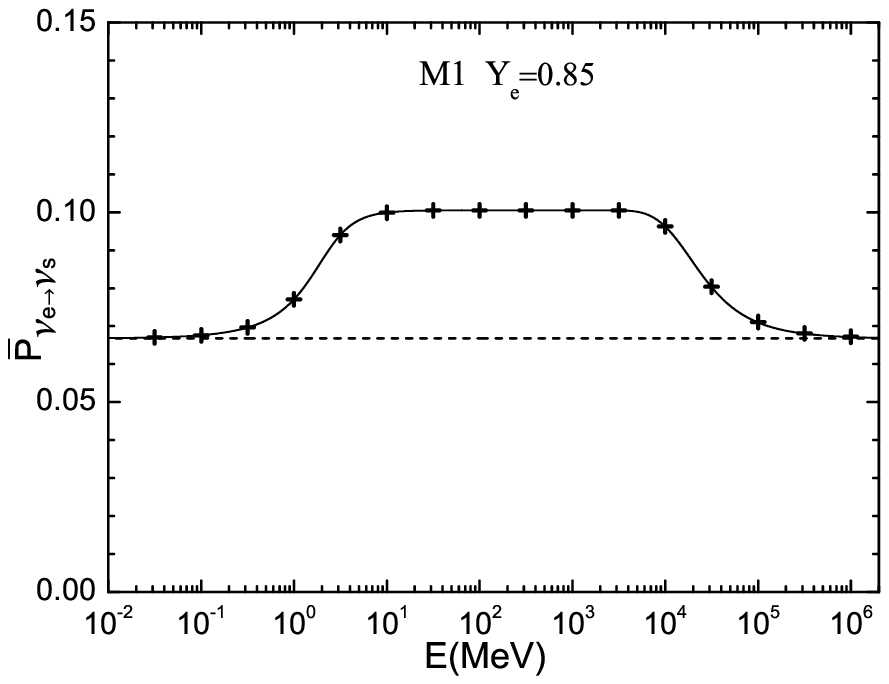}
&
\includegraphics*[bb=18 16 269 208, width=0.46\textwidth]{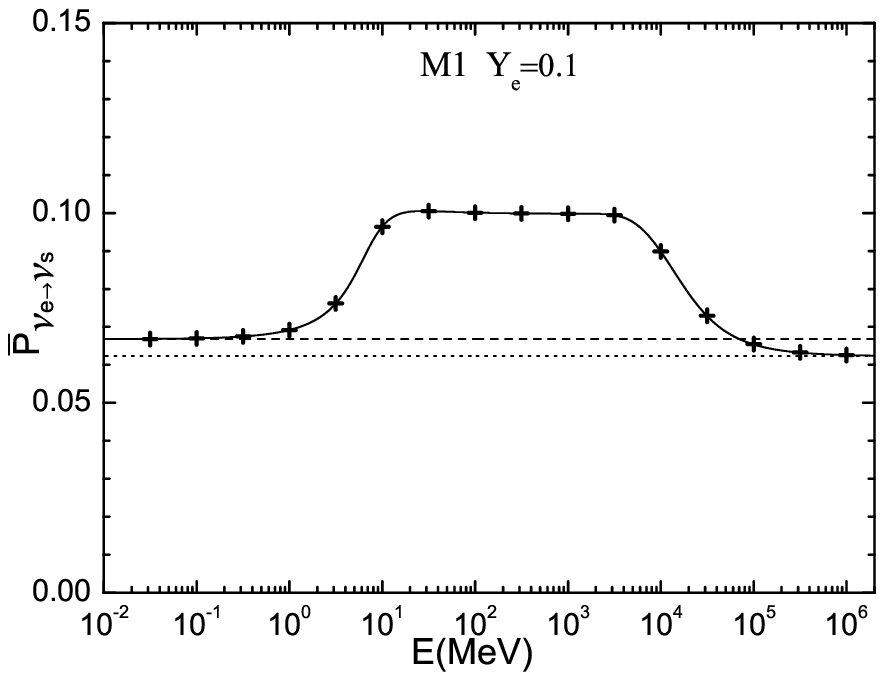}
\\
\includegraphics*[bb=18 16 271 209, width=0.46\textwidth]{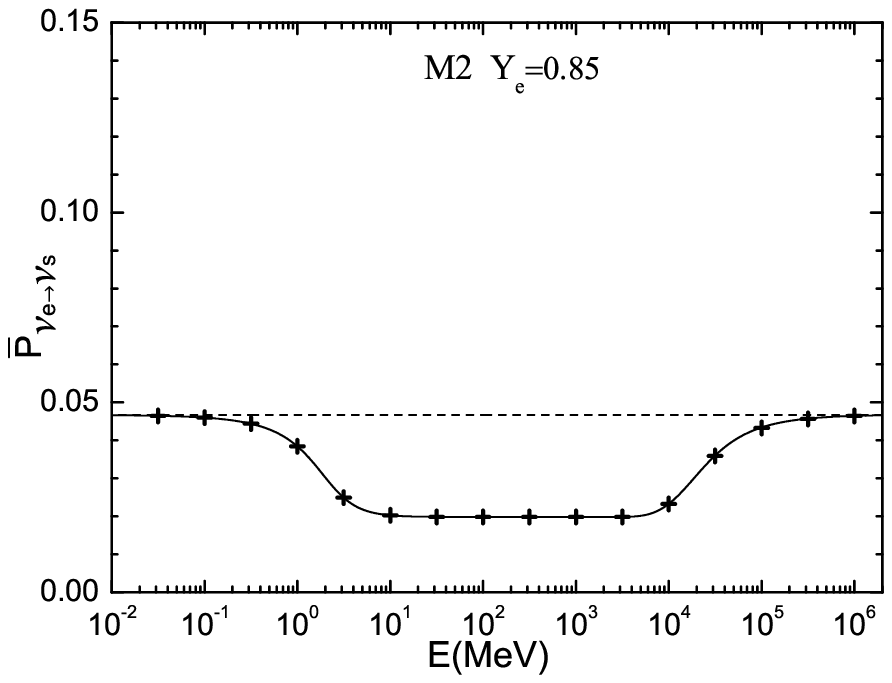}
&
\includegraphics*[bb=18 16 268 209, width=0.46\textwidth]{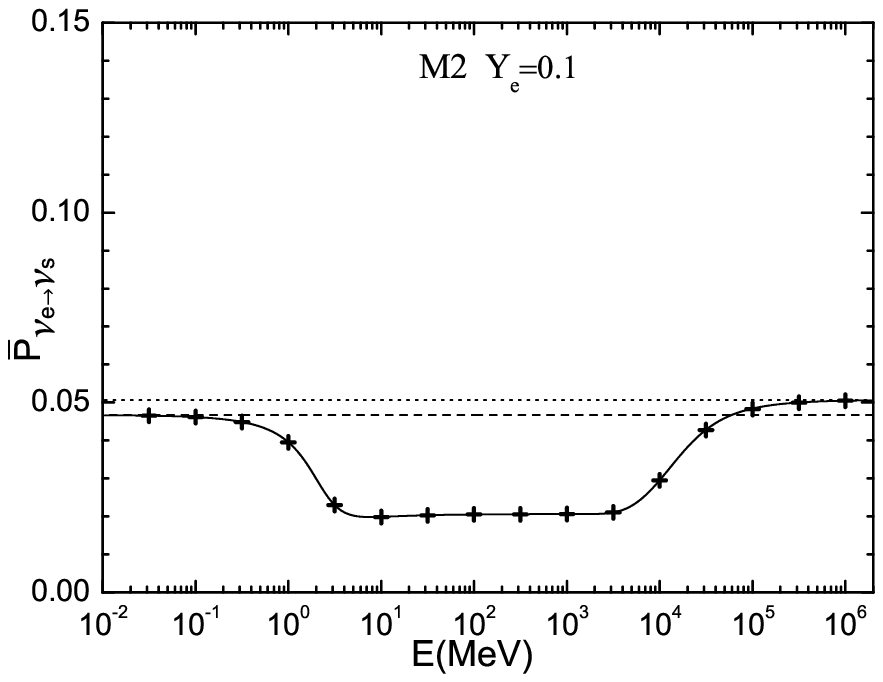}
\\
\includegraphics*[bb=18 16 271 209, width=0.46\textwidth]{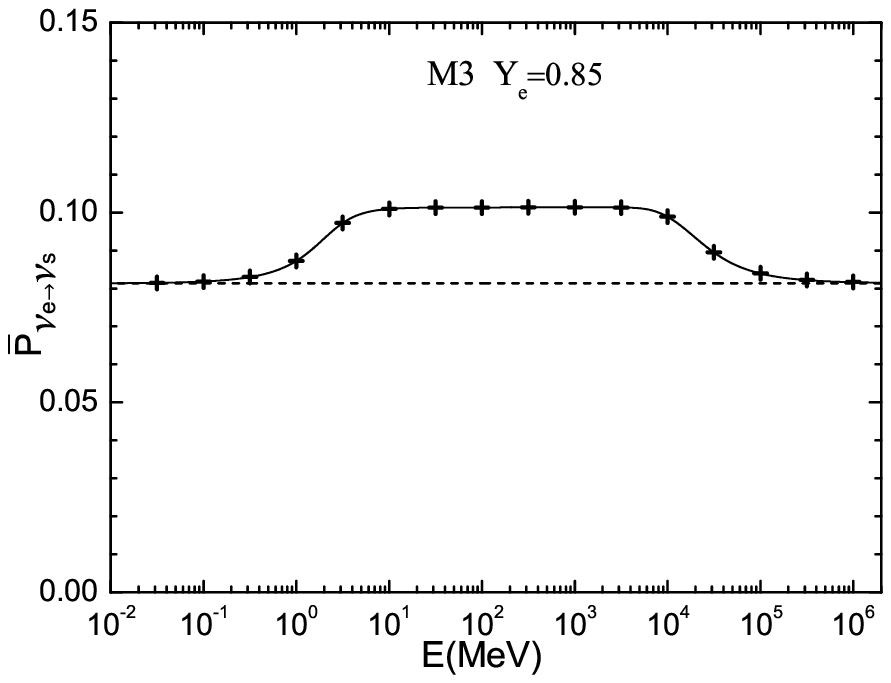}
&
\includegraphics*[bb=18 16 270 209, width=0.46\textwidth]{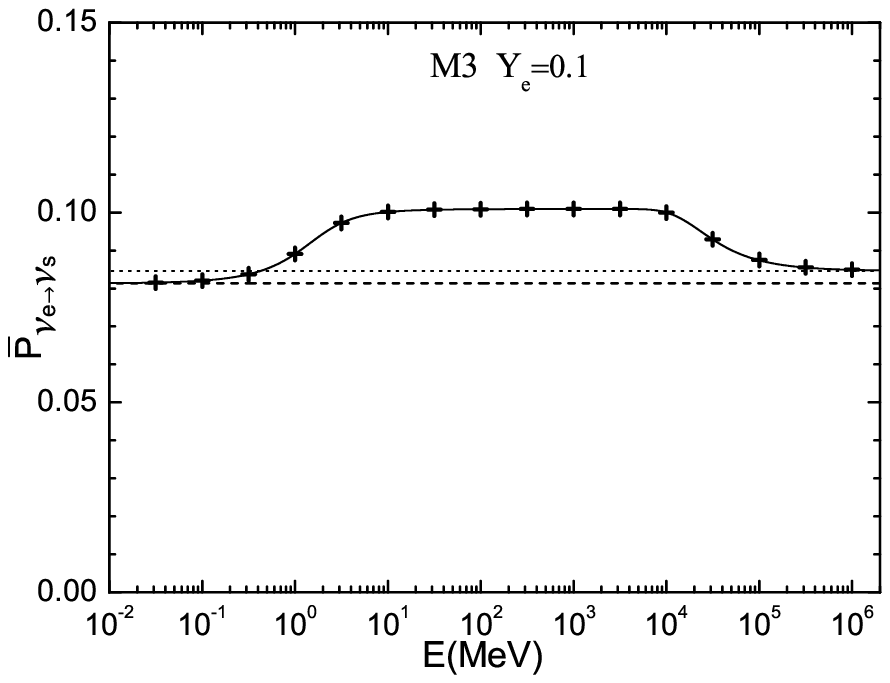}
\end{tabular}
\end{center}
\caption{ \label{fig03}
Averaged probability of $\nu_{e}\to\nu_{s}$ transitions as a function of the neutrino energy $E$
for the three examples of mixing matrices M1, M2, and M3 in Tab.~\ref{tab01}.
For each of them we consider the two values of the electron fraction $Y_{e}$ in Tab.~\ref{tab01}.
In each plot,
the solid line is obtained with the analytic expression in Eq.~(\ref{068}),
the points are obtained with a numerical solution of the evolution equation,
the horizontal dashed line shows the value of $\overline{P}_{\nu_{e}\to\nu_{s}}^{\text{VAC}}$,
and
the horizontal dotted line shows the value of $\overline{P}_{\nu_{e}\to\nu_{s}}^{(\gamma_{\text{R}} \ll 1)}$.
}
\end{figure}

\begin{figure}[t!]
\begin{center}
\begin{tabular}{cc}
\includegraphics*[bb=18 16 272 209, width=0.46\textwidth]{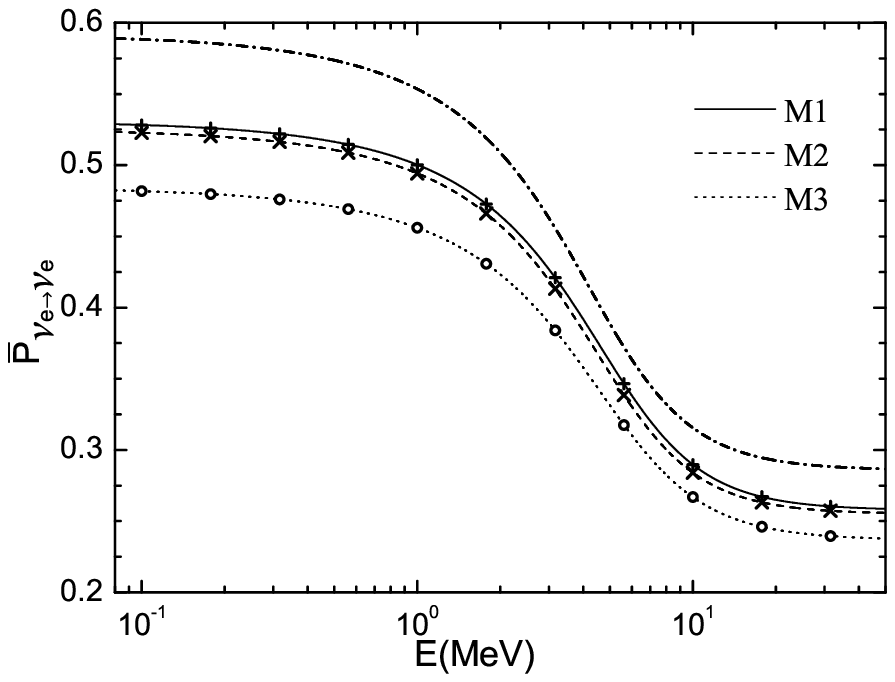}
&
\includegraphics*[bb=18 16 276 208, width=0.46\textwidth]{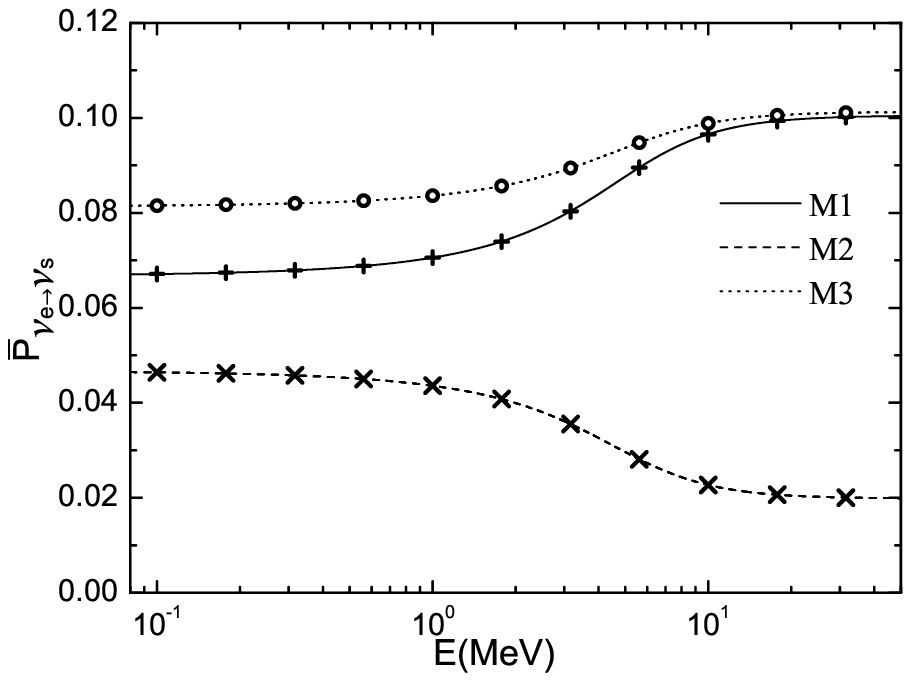}
\end{tabular}
\end{center}
\caption{ \label{fig04}
Averaged probability of $\nu_{e}$ survival and $\nu_{e}\to\nu_{s}$ transitions as functions of the neutrino energy $E$
for the three examples of mixing matrices M1, M2, and M3 in Tab.~\ref{tab01}
calculated for the BP04 Standard Solar Model density \cite{astro-ph/0402114}.
For each of the three examples
the lines are obtained with the analytic expression in Eq.~(\ref{068})
and
the overlapping points are obtained with a numerical solution of the evolution equation.
In the left panel we also plotted the standard three-neutrino mixing value of $\overline{P}_{\nu_{e}\to\nu_{e}}(E)$
in the case of negligible $U_{e3}$ (dash-dotted line).
}
\end{figure}

We solved numerically the effective two-neutrino evolution equation (\ref{041}) for
the exponential electron number density profile
\begin{equation}
N_{e}(x) = 245 \, N_{A} \, \exp\left( - 10.54 \, x / R_{\text{SUN}} \right) \text{cm}^{-3}
\,,
\label{155}
\end{equation}
which is a good approximation of the real electron number density in the solar radiative and convective zones
(see Ref.~\cite{astro-ph/0010346}).
We considered two values of the electron fraction
$Y_{e}$:
the realistic value
$ Y_{e} = 0.85 $,
which is a good approximation in the solar radiative and convective zones
(see Fig.~\ref{fig01}),
and the unrealistic value
$ Y_{e} = 0.1 $,
which is useful to test the effects of the neutral-current potential
and the case $ \xi \neq \vartheta_{e} $.

The panels in Figs.~\ref{fig02} and \ref{fig03} show, respectively,
the averaged $\nu_{e}$ survival probability
and
the averaged probability of $\nu_{e}\to\nu_{s}$ transitions as functions of the neutrino energy $E$
for the three examples of mixing matrices M1, M2, and M3 in Tab.~\ref{tab01}.
Notice that in order to check beyond any doubt the validity of
the analytic flavor transition probability (\ref{068})
both in the adiabatic and non-adiabatic regimes,
we considered a very wide range of energies,
from $ 10^{-2} \, \text{MeV} $ to $ 10^{6} \, \text{MeV} $,
which is much wider than the energy range of measurable solar neutrinos,
from about $ 0.2 \, \text{MeV} $ to about $ 15 \, \text{MeV} $.
Moreover,
the inequality (\ref{026}),
which allows the approximation in Eq.~(\ref{029}),
is not valid for energies not much smaller than about 100 MeV,
for which the value of $A_{\text{CC}}$ in the center of the Sun is
of the same order as
$
|\Delta{m}^{2}_{31}|
\simeq
\Delta{m}^{2}_{\text{ATM}}
$.
Hence, we emphasize that the energy values larger than those of solar neutrinos in
Figs.~\ref{fig02} and \ref{fig03}
are considered only for the purpose of checking the validity of the analytic solution of
the evolution equation (\ref{041}).

From Figs.~\ref{fig02} and \ref{fig03} one can see that
the values of the flavor transition probability
obtained with the analytic expression (\ref{068}) and
the numerical solution of the
evolution equation (\ref{041}) agree perfectly
in all the three examples considered.
Also,
the extreme non-adiabatic limit discussed in Section~\ref{Non-Adiabatic}
is perfectly described by the analytic approximation.

The figures illustrate also some different peculiar behavior of the averaged oscillation probabilities
in the three examples.

With the realistic electron fraction $ Y_{e} = 0.85 $,
$\overline{P}_{\nu_{e}\to\nu_{e}}$
has the same qualitative behavior in the three examples,
whereas
$\overline{P}_{\nu_{e}\to\nu_{s}}$
is enhanced by the matter effect in M1 and M3
and suppressed in M2.
This is due to the different signs of 
$\cos2\vartheta_{s}$
(see Tab.~\ref{tab01}).
In M1 and M3 $\cos2\vartheta_{s}$ is negative
and the adiabatic part of the transition probability
$ 0.5 \cos2\vartheta_{s} \cos2\vartheta_{e}^{0} \simeq - 0.5 \cos2\vartheta_{s} $
is positive, leading to an enhancement of $\overline{P}_{\nu_{e}\to\nu_{s}}$.
In M2 the opposite happens.

In the case of the unrealistic electron fraction $ Y_{e} = 0.1 $,
one can see from Figs.~\ref{fig02} and \ref{fig03}
that 
$ \overline{P}_{\nu_{e}\to\nu_{e}}^{(\gamma_{\text{R}} \ll 1)} \neq \overline{P}_{\nu_{e}\to\nu_{e}}^{\text{VAC}} $
because
$ \xi \neq \vartheta_{e} $,
as discussed in Section~\ref{Non-Adiabatic}.
The three examples have the following interesting differences in the extreme non-adiabatic limit.
From Fig.~\ref{fig02} one can see that
$ \overline{P}_{\nu_{e}\to\nu_{e}}^{(\gamma_{\text{R}} \ll 1)} > \overline{P}_{\nu_{e}\to\nu_{e}}^{\text{VAC}} $
in M1 and M2
and the inequality is reversed in M3.
This happens because
$ \xi < \vartheta_{e} $ in M1 and M2,
whereas
$ \xi > \vartheta_{e} $ in M3.
From Fig.~\ref{fig03} one can see that
$ \overline{P}_{\nu_{e}\to\nu_{s}}^{(\gamma_{\text{R}} \ll 1)} < \overline{P}_{\nu_{e}\to\nu_{s}}^{\text{VAC}} $
in M1
and the inequality is reversed in M2 and M3.
These behaviors are due to the fact that
$ \xi < \vartheta_{e} $ and $\cos2\vartheta_{s}<0$ in M1,
$ \xi < \vartheta_{e} $ and $\cos2\vartheta_{s}>0$ in M2,
and
$ \xi > \vartheta_{e} $ and $\cos2\vartheta_{s}<0$ in M3.

After the check of the analytic expression (\ref{068}) for the flavor transition probability
in the case of the perfect exponential density (\ref{155}) and a constant electron number fraction,
we also checked its validity for the realistic BP04 Standard Solar Model density \cite{astro-ph/0402114},
taking into account the variation of the electron number fraction (plotted in Fig.~\ref{fig01}).
The results of the analytic expression (\ref{068}) and of the numerical solution of the evolution equation (\ref{041})
are presented in Fig.~\ref{fig04} for the three examples of mixing matrices M1, M2, and M3.
We considered only neutrino energies smaller than about 50 MeV, for which the inequality (\ref{026}) and the approximation (\ref{029}) are valid.
In this range of energies the flavor transition is practically adiabatic and we expect that
non-adiabatic effects due to the variation of the electron number fraction are negligible.
Indeed,
from Fig.~\ref{fig04}
one can see that the analytic approximation is very accurate for solar neutrinos and the
effects of the variation of the electron number fraction,
which are neglected in the analytic approximation, are really negligible.

From the left panel in Fig.~\ref{fig04} one can see that active-sterile transitions as those in the examples M1, M2, and M3
can change significantly the $\nu_{e}$ survival probability with respect to the standard
three-neutrino mixing value in the case of negligible $U_{e3}$ (dash-dotted line).
Therefore,
it is important to perform an analysis of solar neutrino data
\cite{CG-ML-YFL-QYL-2009}
in the framework of active-sterile neutrino mixing that we have considered
in order to obtain information on the possible existence of sterile neutrinos and their mixing with the active ones.

\section{Conclusions}
\label{092}
\nopagebreak

In this paper we derived an analytic solution for the flavor transition probabilities
of solar neutrinos in a general scheme with an arbitrary number of sterile neutrinos,
without any constraint on the mixing.
Hence, we improved the study presented in Ref.~\cite{hep-ph/9908513},
where it was assumed that the electron neutrino has non-negligible mixing only with the two massive neutrinos
which generate the solar squared-mass difference.
Here we only assumed a realistic hierarchy of neutrino squared-mass differences
in which the smallest squared-mass difference is effective in solar neutrino oscillations.

In Section~\ref{091} we have illustrated the validity of the analytical approximation
(\ref{068})
of the flavor transition probability
by comparing its value with that obtained with a numerical solution of the evolution equation
in three examples of the possible mixing matrix in the simplest case of four-neutrino mixing.
Hence, we are confident that the expression (\ref{068})
of the flavor transition probability
can be used in a reliable analysis of the solar neutrino data
which could provide important information on
active-sterile neutrino mixing
\cite{CG-ML-YFL-QYL-2009}.

\section*{Acknowledgments}
\nopagebreak
We would like to thank the Department of Theoretical Physics of the University of Torino
for hospitality and support.

\appendix

\section{Relations Between Bases}
\label{appendix}

In this paper we used four different bases of the neutrino states and amplitudes:

\begin{enumerate}

\item
The flavor basis of amplitudes in Eq.~(\ref{017}),
related to the flavor states on the left-hand side of Eq.~(\ref{005a}).

\item
The vacuum mass basis of amplitudes in Eq.~(\ref{027}),
related to the vacuum mass basis of states in the right-hand side of Eq.~(\ref{005a}).

\item
The effective mass basis in matter of amplitudes in the truncated 1-2 sector,
$\psi^{\text{M}}_{1}$ and $\psi^{\text{M}}_{2}$,
introduced in Eq.~(\ref{042}).

\item
The effective interaction basis in matter of amplitudes in the truncated 1-2 sector,
$\psi^{\text{I}}_{1}$ and $\psi^{\text{I}}_{2}$,
introduced in Eq.~(\ref{044}).

\end{enumerate}

In this Appendix we derive the relations between the effective mass and interaction bases in matter and the flavor basis of the amplitudes and
the relations of the corresponding states.

Since
\begin{equation}
\begin{pmatrix}
\psi^{\text{V}}_{1}
\\
\psi^{\text{V}}_{2}
\end{pmatrix}
=
\begin{pmatrix}
U_{e1}^{*}
&
U_{\mu1}^{*}
&
U_{\tau1}^{*}
&
U_{s_{1}1}^{*}
&
\cdots
\\
U_{e2}^{*}
&
U_{\mu2}^{*}
&
U_{\tau2}^{*}
&
U_{s_{1}2}^{*}
&
\cdots
\end{pmatrix}
\begin{pmatrix}
\psi_{e}
\\
\psi_{\mu}
\\
\psi_{\tau}
\\
\psi_{s_{1}}
\\
\vdots
\end{pmatrix}
\,,
\label{046}
\end{equation}
we have
\begin{equation}
\begin{pmatrix}
\psi^{\text{I}}_{1}
\\
\psi^{\text{I}}_{2}
\end{pmatrix}
=
\begin{pmatrix}
\cos\xi U_{e1}^{*} + \sin\xi U_{e2}^{*}
&
\cos\xi U_{\mu1}^{*} + \sin\xi U_{\mu2}^{*}
&
\cdots
\\
- \sin\xi U_{e1}^{*} + \cos\xi U_{e2}^{*}
&
- \sin\xi U_{\mu1}^{*} + \cos\xi U_{\mu2}^{*}
&
\cdots
\end{pmatrix}
\begin{pmatrix}
\psi_{e}
\\
\psi_{\mu}
\\
\vdots
\end{pmatrix}
\,.
\label{047}
\end{equation}
Therefore,
the effective interaction amplitudes $\psi^{\text{I}}_{1}$ and $\psi^{\text{I}}_{2}$
for which the matter potential
in the truncated 1-2 sector is diagonal
are linear combinations of the flavor amplitudes
$\psi_{e}$, $\psi_{\mu}$, $\psi_{\tau}$, $\psi_{s_{1}}$, $\ldots$, $\psi_{s_{N_{s}}}$.
Furthermore,
from Eqs.~(\ref{042}) and (\ref{046}) we derive immediately the relation between
the amplitudes $\psi^{\text{M}}_{1}$ and $\psi^{\text{M}}_{2}$ of the effective energy eigenstates in matter
and the flavor amplitudes
$\psi_{e},\psi_{\mu},\psi_{\tau},\psi_{s_{1}},\ldots,\psi_{s_{N_{s}}}$:
\begin{equation}
\begin{pmatrix}
\psi^{\text{M}}_{1}
\\
\psi^{\text{M}}_{2}
\end{pmatrix}
=
\begin{pmatrix}
\cos\omega U_{e1}^{*} - \sin\omega U_{e2}^{*}
&
\cos\omega U_{\mu1}^{*} - \sin\omega U_{\mu2}^{*}
&
\cdots
\\
\sin\omega U_{e1}^{*} + \cos\omega U_{e2}^{*}
&
\sin\omega U_{\mu1}^{*} + \cos\omega U_{\mu2}^{*}
&
\cdots
\end{pmatrix}
\begin{pmatrix}
\psi_{e}
\\
\psi_{\mu}
\\
\vdots
\end{pmatrix}
\,.
\label{048}
\end{equation}

In the effective mass and interaction bases in matter
the solar neutrino state in Eq.~(\ref{014}) is given by
\begin{align}
| \nu(x) \rangle
=
\null & \null
\sum_{k=1}^{2} \psi^{\text{M}}_{k}(x) | \nu^{\text{M}}_{k} \rangle
+
\sum_{k=3}^{N} \psi^{\text{V}}_{k}(x) | \nu^{\text{V}}_{k} \rangle
\nonumber
\\
=
\null & \null
\sum_{k=1}^{2} \psi^{\text{I}}_{k}(x) | \nu^{\text{I}}_{k} \rangle
+
\sum_{k=3}^{N} \psi^{\text{V}}_{k}(x) | \nu^{\text{V}}_{k} \rangle
\,,
\label{014b}
\end{align}
with the effective massive states in matter
\begin{align}
\begin{pmatrix}
| \nu^{\text{M}}_{1} \rangle
\\
| \nu^{\text{M}}_{2} \rangle
\end{pmatrix}
=
\null & \null
\begin{pmatrix}
\cos\omega
&
- \sin\omega
\\
\sin\omega
&
\cos\omega
\end{pmatrix}
\begin{pmatrix}
| \nu^{\text{V}}_{1} \rangle
\\
| \nu^{\text{V}}_{2} \rangle
\end{pmatrix}
\nonumber
\\
=
\null & \null
\begin{pmatrix}
\cos\omega U_{e1} - \sin\omega U_{e2}
&
\cos\omega U_{\mu1} - \sin\omega U_{\mu2}
&
\cdots
\\
\sin\omega U_{e1} + \cos\omega U_{e2}
&
\sin\omega U_{\mu1} + \cos\omega U_{\mu2}
&
\cdots
\end{pmatrix}
\begin{pmatrix}
| \nu_{e} \rangle
\\
| \nu_{\mu} \rangle
\\
\vdots
\end{pmatrix}
\,,
\label{085}
\end{align}
and the effective interaction states
\begin{align}
\begin{pmatrix}
| \nu^{\text{I}}_{1} \rangle
\\
| \nu^{\text{I}}_{2} \rangle
\end{pmatrix}
=
\null & \null
\begin{pmatrix}
\cos\xi
&
\sin\xi
\\
- \sin\xi
&
\cos\xi
\end{pmatrix}
\begin{pmatrix}
| \nu^{\text{V}}_{1} \rangle
\\
| \nu^{\text{V}}_{2} \rangle
\end{pmatrix}
=
\begin{pmatrix}
\cos\xi_{\text{M}}
&
\sin\xi_{\text{M}}
\\
- \sin\xi_{\text{M}}
&
\cos\xi_{\text{M}}
\end{pmatrix}
\begin{pmatrix}
| \nu^{\text{M}}_{1} \rangle
\\
| \nu^{\text{M}}_{2} \rangle
\end{pmatrix}
\nonumber
\\
=
\null & \null
\begin{pmatrix}
\cos\xi U_{e1} + \sin\xi U_{e2}
&
\cos\xi U_{\mu1} + \sin\xi U_{\mu2}
&
\cdots
\\
- \sin\xi U_{e1} + \cos\xi U_{e2}
&
- \sin\xi U_{\mu1} + \cos\xi U_{\mu2}
&
\cdots
\end{pmatrix}
\begin{pmatrix}
| \nu_{e} \rangle
\\
| \nu_{\mu} \rangle
\\
\vdots
\end{pmatrix}
\,.
\label{086}
\end{align}
Thus,
a flavor neutrino states can be expressed as
\begin{align}
| \nu_{\alpha} \rangle
=
\null & \null
\sum_{k=1}^{N} U_{\alpha k}^{*} | \nu^{\text{V}}_{k} \rangle
\nonumber
\\
=
\null & \null
\left( \cos\omega U_{\alpha1}^{*} - \sin\omega U_{\alpha2}^{*} \right) | \nu^{\text{M}}_{1} \rangle
+
\left( \sin\omega U_{\alpha1}^{*} + \cos\omega U_{\alpha2}^{*} \right) | \nu^{\text{M}}_{2} \rangle
+
\sum_{k=3}^{N} U_{\alpha k}^{*} | \nu^{\text{V}}_{k} \rangle
\nonumber
\\
=
\null & \null
\left( \cos\xi U_{\alpha1}^{*} + \sin\xi U_{\alpha2}^{*} \right) | \nu^{\text{I}}_{1} \rangle
+
\left( - \sin\xi U_{\alpha1}^{*} + \cos\xi U_{\alpha2}^{*} \right) | \nu^{\text{I}}_{2} \rangle
+
\sum_{k=3}^{N} U_{\alpha k}^{*} | \nu^{\text{V}}_{k} \rangle
\,,
\label{086a}
\end{align}
for
$\alpha=e,\mu,\tau,s_{1},\ldots,s_{N_{s}}$.
One can see that in general each flavor state contains
both the effective massive states
$| \nu^{\text{M}}_{1} \rangle$
and
$| \nu^{\text{M}}_{2} \rangle$
as well as
both the effective interaction states
$| \nu^{\text{I}}_{1} \rangle$
and
$| \nu^{\text{I}}_{2} \rangle$.

\raggedright


\end{document}